\documentclass[12pt,noshowpacs,nofootinbib,notitlepage,amsmath]{revtex4-1}
\usepackage{setspace}
\allowdisplaybreaks
\usepackage{graphicx,color}
\usepackage[colorlinks=true,citecolor=blue,linkcolor=blue,urlcolor=blue]{hyperref}
\usepackage[charter]{mathdesign}
\usepackage{multirow}
\usepackage{enumerate}
\usepackage{amsmath}
\renewcommand{\arraystretch}{2}

\usepackage{array}
\usepackage{booktabs} 

\DeclareSymbolFontAlphabet{\mathcal}{symbols}
\DeclareSymbolFont{symbols}{OMS}{xmdcmsy}{m}{n}
\DeclareSymbolFont{largesymbols}{OMX}{xmdcmex}{m}{n}
\SetSymbolFont{symbols}{bold}{OMS}{xmdcmsy}{b}{n}

\def\Mp{m_{\mathrm{Pl}}}
\def\lp{\ell_{\mathrm{Pl}}}

\begin{document}  
\title{\color{blue}\Large Anatomy of a thermal black hole mimicker}
\author{Jing Ren}
\email{renjing@ihep.ac.cn}
\affiliation{Institute of High Energy Physics, Chinese Academy of Sciences,\\
Beijing 100049, China}

\begin{abstract}
We are entering a new era to test the strong gravity regime around astrophysical black holes. The possibility that they are actually horizonless ultracompact objects and then free from the information loss paradox can be examined more closely with observational data. In this paper, we systematically develop a thermal gas model of the 2-2-hole in quadratic gravity, as one step further to look for more tractable models of black hole mimickers. Concrete predictions for departures from black holes are made all the way down to the high curvature interior. The simple form of matter further enables an explicit study of the relation between geometry and thermodynamics. Within this unified framework, we identify notably different behaviors at two limits. On one side is the astrophysically large 2-2-hole, as characterized by a minuscule deviation outside the would-be horizon and a highly squeezed interior along the radial direction. Anomalous features of black hole thermodynamics emerge from the ordinary gas. On the other side is the minimal 2-2-hole with an isotropic and shrinking interior, which behaves more like a normal thermodynamic system. This brings a new perspective to the related theoretical questions as well as phenomenological implications.
\end{abstract}
\maketitle

\section{Introduction}
\label{intro}

Until very recently, we learned about the strong gravity regime around astrophysical black holes mainly through extrapolation from the observation at rather large distance, given that ultracompact objects in general relativity (GR) can only be black holes as described by the Kerr-Newman metric. 
The detections of gravitational waves from compact binary mergers for stellar-mass black holes~\cite{LIGOScientific:2019fpa} and the first image of a supermassive black hole with the shadow~\cite{Akiyama:2019cqa} open new windows for the exploration at horizon scales. 
The consistency of current measurements with GR predictions pushes the constraint down to the photon sphere outside the horizon, while little can be said about the plausible departures at a much closer distance. 

Theoretically near horizon modifications are expected to resolve the long-standing black hole information loss paradox~\cite{tHooft:2016fzb}. One extreme possibility is that the horizon formation might be halted due to some new physics, and gravitational collapse actually ends up with a horizonless ultracompact object that deviates from a black hole only at a minuscule distance outside the would-be horizon. Many candidates are proposed along this line, such as fuzzballs~\cite{Mathur:2005zp} and gravastars~\cite{Mazur:2004fk}.
Some of them make intriguing connections to quantum gravity.
\footnote{Macroscopic quantum gravity effects at the wound-be horizon are also possible, as discussed in \cite{Dvali:2011aa,Giddings:2012bm}.}
However, given the large hierarchy between the Planck scale and the curvature scale around a macroscopic horizon, quantum gravity effects are normally not expected at the horizon scales. 
Not much insight has been provided for this question yet. Moreover, well-motivated and concrete predictions for departures from black holes are still lacking. 
The new era of observational astronomy provides us a great chance to  look for the related new phenomena, e.g. gravitational wave echoes~\cite{Cardoso:2016oxy}, tidal love number~\cite{Cardoso:2017cfl} and so on. Developing more tractable models for horizonless ultracompact objects is then strongly motivated. 
See \cite{Cardoso:2019rvt} for a status report on horizonless ultracompact objects, including relevant observational constraints and existing theoretical models.

Quantum quadratic gravity was known to be a renormalizable and asymptotically free quantum field theory of gravity for decades~\cite{Stelle:1976gc, Voronov:1984kq, Fradkin:1981iu, Avramidi:1985ki}. The notorious ghost problem as related to the higher derivative terms nonetheless prevents it to be accepted as a UV completion to GR. Although the critique based on the classical picture seems quite convincing, quantum mechanism might be crucial for the final words.\footnote{In the path integral formulation of quantum field theory, the measure could be as important as the classical Hamiltonian, and the physical spectrum shall be determined by the dressed-propagator.} There is an ongoing effort to resolve the problem by taking quantum corrections seriously. Solutions depend on the strength of running gravitational couplings for quadratic curvature terms at the mass scale of classical modes. If couplings remain weak, the ghost pole may be removed systematically by modifying the quantum prescription~\cite{Lee:1969fy, Tomboulis:1977jk, Grinstein:2008bg, Anselmi:2017yux, Donoghue:2018lmc} or the probability interpretation~\cite{Bender:2007wu, Salvio:2015gsi}. If couplings already become strong at some higher scale, say for a pure quadratic action defined at the UV, there is the possibility that the strong gravity dynamically generates the Planck scale and removes the would-be ghost simultaneously~\cite{Holdom:2015kbf, Holdom:2016xfn}. More discussions of the theory can be found in~\cite{Salvio:2018crh, Holdom:2019}.

In comparison to other candidates of quantum gravity, quadratic gravity provides a weakly coupled field theory description for gravity at the high energy scale, and so a more tractable framework to study high curvature effects around macroscopic black holes.
To be specific, a classical action (classical quadratic gravity) is used to find nontrivial background solutions as approximations to configurations in the quantum theory,
\begin{eqnarray}\label{eq:CQG}
S_{\mathrm{CQG}}=\frac{1}{16\pi}\int d^4x\,\sqrt{-g}\left(\Mp^2R-\alpha C_{\mu\nu\alpha\beta}C^{\mu\nu\alpha\beta}+\beta R^2\right).
\end{eqnarray}
The dimensionless couplings $\alpha, \beta$ determine the mass scale $m_2=\Mp/\sqrt{2\alpha}$ and $m_0=\Mp/\sqrt{6\beta}$ for the new spin-2 and spin-0 modes respectively. 
In the weak couplings scenario, where $\alpha, \beta\gg 1$ and $m_i \ll \Mp$,
eq.(\ref{eq:CQG}) shows the action when the Planck mass has been generated by the scalar vacuum expectation value~\cite{Salvio:2014soa}. It is also a good approximation at super-Planckian curvature. A smaller upper bound $m_i\lesssim 1$\,TeV is further motivated by the Higgs boson hierarchy problem. 
To not ruin the precision test of GR in the solar system, it is safe to have $m_i\gtrsim 10^{-10}$\,eV, for which the Compton wavelength is no larger than $\mathcal{O}(km)$.
In the strong coupling scenario, there is only one mass scale in the theory, so we have $\alpha, \beta\sim \mathcal{O}(1)$ and $m_i\sim \Mp$. The classical action (\ref{eq:CQG}) has the same limit as the quantum theory for both small and large curvatures~\cite{Holdom:2015kbf, Holdom:2016xfn,Holdom:2016nek}. \footnote{The logarithmic running of dimensionless couplings at high energy is a subleading effect and ignored here.}

As been found in our earlier work~\cite{Holdom:2016nek}, a compact matter distribution doesn't necessarily lead to the formation of horizon in classical quadratic gravity (\ref{eq:CQG}).
Actually, a new family of horizonless solution appears when matter distribution shrinks within the would-be horizon $r_H$. We called this new type of solutions the 2-2-hole, given that the metric vanishes as $r^2$ when approaching the origin.
\footnote{The black hole solutions still exist~\cite{Lu:2015psa}, which go to the GR limit when the quadratic curvature terms are turned off. Here we focus on the new family of horizonless solution that has no analogue in GR.}
Focusing on the strong gravity scenario~\cite{Holdom:2016nek},
we found that an astrophysical 2-2-hole closely resembles the Schwarzschild metric at large distance with exponentially small corrections from new massive modes. Drastic deviations occur at about Planck distance outside the would-be horizon, as determined by the unique scale $m_i\sim\Mp$ in the action. Inside a narrow transition region, the metric quickly approaches the characteristic $r^2$ behavior without changing the sign. This implies a rather deep gravitational potential for the interior, the radial proper length of which shrinks to be only the order of Planck length. The curvatures also quickly become super-Planckian and reach a curvature singularity at the origin. 
This may cause the geodesic incompleteness problem for point particles. 
But given the quantum nature of particles, waves with finite energy might be more appropriate to consider as probes for such extreme conditions.  
We found that the 2-2-hole time-like singularity can appear regular for the wave. With a unique boundary condition automatically imposed at the origin, there is no ambiguity of its time evolution. This boundary condition is furthermore of a perfectly reflecting type and plays an essential role for generating gravitational wave echoes and for the microscopic state counting~\cite{Holdom:2016nek,  Holdom:2019}.
As a candidate for horizonless ultracompact objects, the 2-2-hole then stands out for two reasons. Instead of what the naive dimensional analysis suggests, microscopic deviations around a macroscopic would-be horizon are shown to be possible, as driven by the new quadratic curvature terms.
Also it features a novel interior as compared to a star.  

Our previous study found 2-2-holes sourced by a thin-shell of matter with exotic equation of state at a small shell radius $\ell\lesssim r_H$. This helps identify crucial properties of 2-2-holes as described above, but less can be learned for the relation between geometry and matter properties. 
If 2-2-holes serve as a general description for the endpoint of gravitational collapse, it shall not depend that much on the explicit form of matter. 
Moreover, with the high curvature inside, infalling matter shall be easily disrupted by the tidal force. The gas particles are further thermalized and reach equilibrium after a certain time. 
Recently 2-2-holes sourced by the relativistic thermal gas have been found numerically~\cite{Holdom:2019}. 
This provides an interesting physical model of 2-2-holes with a quite simple matter source. In particular, it enables the study of 2-2-hole thermodynamics as an ordinary system, similar to that for self-gravitating radiation in GR~\cite{Sorkin:1981wd}. 
The difference is that now radiation is able to support an ultracompact configuration in equilibrium without collapsing into a black hole due to the extra quadratic curvature terms.
It is then natural to make comparison with black hole thermodynamics, the origin of which is still mysterious and is thought to be closely related to the information loss paradox~\cite{Harlow:2014yka, Polchinski:2016hrw}.   

In this paper, we systematically study the thermal gas model for 2-2-holes.     
The Weyl tensor term $C_{\mu\nu\alpha\beta}C^{\mu\nu\alpha\beta}$ turns out to be essential for the existence of 2-2-holes, while the $R^2$ term plays little role. So we focus on 2-2-holes in the Einstein-Weyl theory with $\beta=0$ in (\ref{eq:CQG}). The spin-2 mass $m_2$ (or its Compton wavelength $\lambda_2\equiv 1/m_2$) is then the only free parameter, which we allow to vary in a wide range to account for both the weak and strong coupling scenarios of the quantum theory. 
The thermal gas in equilibrium entails a quite compact matter distribution and can always source a 2-2-hole. The solution is determined by the relative difference between the object's size $r_H$ and the Compton wavelength $\lambda_2$. A 2-2-hole always exists in the large mass limit $r_H/\lambda_2\gg 1$. It is characterized by a quite narrow transition region and a highly curved interior squeezed along the radial direction as described above. Some intriguing features of black hole thermodynamics can be reproduced in this limit.
When $r_H$ becomes comparable to $\lambda_2$, the 2-2-hole has a much broader transition region and a rapidly shrinking interior. The resemblance to black hole thermodynamics ceases to apply. The solution no longer exists when $r_H/\lambda_2\lesssim1$, as we might expect from the uncertainty principle.
This paints a more complete picture for 2-2-holes, with rich implications for phenomenology. 

The rest of the paper is organized as follows. The 2-2-hole sourced by the relativistic thermal gas is studied in Sec.~\ref{sec:thegas}, where we thoroughly explore the solution space by keeping the full dependence on all relevant scales. Their properties in the large and small mass limits are discussed in Sec.~\ref{sec:largeM} and Sec.~\ref{sec:smallM} respectively, including the interior scaling behaviors for the metric and thermodynamic variables, and also the physical implications. In Sec.~\ref{sec:variation} we study some variations of the relativistic thermal gas model. 
A generalization to the thermal gas with nonzero particle mass is discussed in Sec.~\ref{sec:thegas2}, where we check how thermodynamic variables change with matter properties.
Sec.~\ref{sec:shell} considers a 2-2-hole perturbed by a matter shell. It serves as a toy model to see how 2-2-holes grow slowly with matter accretion. We conclude in Sec.~\ref{sec:conc}. 
The field equations in the Einstien-Weyl theory are summarized in Appendix.~\ref{sec:appA}. The series expansions for 2-2-holes in the thermal gas model are presented in Appendix.~\ref{sec:appB}.

\section{Relativistic thermal gas model}
\label{sec:thegas}

The endpoint of gravitational collapse in quadratic gravity could be a 2-2-hole filled with hot gas  particles in thermal equilibrium. For a static and spherically symmetric spacetime, we can always choose a coordinate system with the following line element
\begin{eqnarray}\label{eq:ds2}
ds^2=-B(r)dt^2+A(r)dr^2+r^2 d\Omega^2\,.
\end{eqnarray} 
The metric functions $A(r), B(r)$ are determined by two field equations, i.e. (\ref{eq:EoM1}) in the Einstein-Weyl gravity. In this coordinate system, a 2-2-hole is defined by the following characteristic leading order behavior of the metric under the series expansion around the origin,  
\begin{eqnarray}
A(r)=a_2r^2+...,\quad
B(r)=b_2r^2+...\,.
\end{eqnarray}
The metric is vanishing at the origin and it has no analog in GR.

The stress tensor of a thermal gas is described by the proper energy density and the isotropic pressure,  
\begin{eqnarray}\label{eq:stresstensor}
T_{\mu\nu}&=&\textrm{diag}\,\left(B \rho, \,A\, p, \,r^2 p, \,r^2 s_\theta^2 p\right),\nonumber\\
\rho&=&\frac{N}{(2\pi)^3}\int_0^\infty \frac{E}{e^{E/T}-\epsilon}4\pi p^2 dp,\nonumber\\
p&=&\frac{N}{3(2\pi)^3}\int_0^\infty \frac{p^2/E}{e^{E/T}-\epsilon}4\pi p^2 dp,
\end{eqnarray}
where $E^2=p^2+m^2$ and $T$ is the locally measured temperature. $N$ denotes the number of particle species. $\epsilon=\pm1$ is for boson and fermion respectively. The total energy and entropy of the thermal gas are,\footnote{As mentioned in Sec.~\ref{intro}, the quantum nature of particles shall be taken into account, and they are neither emitted nor absorbed by the singularity~\cite{Holdom:2016nek,  Holdom:2019}.}
\begin{eqnarray}\label{eq:entropy}
U&=&\int dr \sqrt{A(r) B(r)} \,4\pi r^2 \rho(r), \nonumber\\
S&=&\int dr \sqrt{A(r)} \, 4\pi r^2 s(r),\quad s(r)=\frac{\rho(r)+p(r)}{T(r)}\,.
\end{eqnarray} 
The stress tensor has to satisfy the momentum conservation law: $\nabla^\mu T_{\mu r}=0$. For (\ref{eq:stresstensor}), it is 
\begin{eqnarray}\label{eq:consL1}
p'+(p+\rho)\frac{B'}{2B}=0\,,
\end{eqnarray}
where the derivative is with respect to $r$. 

In this section we focus on the relativistic limit, where the thermal gas particle mass $m$ is negligible. The stress tensor is then traceless $T_\mu^\mu=0$, with the simple equation of state, 
\begin{eqnarray}\label{eq:rhop1}
\rho=3p=N \frac{3}{\pi^2} T^4,\quad
s=N \frac{4}{\pi^2} T^3\,.
\end{eqnarray} 
For simplicity we use the numerical coefficients for $\epsilon=0$ here and absorb the small difference for $\epsilon=\pm1$ cases into $N$. 
The conservation law can be solved analytically, and the matter profile is determined up to a constant with $p(r) B(r)^2=p_\infty$, where $p_\infty=N T_\infty^4/\pi^2$. This corresponds to the Tolman's law for the local temperature $T(r)B(r)^{1/2}=T_\infty$. 
The total entropy and energy of the relativistic thermal gas are then, 
\begin{eqnarray}\label{eq:SUm0}
S=\frac{4}{3}\frac{U}{T_\infty}
=N \frac{16}{\pi}T_\infty^3\int dr  \sqrt{\frac{A(r)}{B(r)^3}}r^2
\,. 
\end{eqnarray} 
As we will see below, the gas outside a 2-2-hole can be quite thin and cold. 
It then will be easily driven away from the equilibrium due to interaction with the environment. For such case,  $T_\infty$ can be viewed as the temperature measured by an observer near infinity.

The field equations are greatly simplified for a traceless stress tensor. Substituting the conservation law, (\ref{eq:EoM1}) becomes 
\begin{eqnarray}\label{eq:EoM2}
H_1=0,\quad H_2=8\pi \frac{A }{B^2} p_\infty\,.
\end{eqnarray}
Without the Birkhoff's theorem, the 2-2-hole solution can only be found numerically. Here the relevant scales are: the Planck length $\lp$, the Compton wavelength for the new spin-2 mode $\lambda_2=1/m_2$, the would-be horizon size $r_H=2M\lp^2$ ($M$ is the physical mass), a new emerging scale related to the interior $r_a\equiv 1/\sqrt{a_2}$, and the source property $p_\infty$. 
The numerical solutions are found in the following way. For a given pair of $(\lambda_2, r_a)$, we do shooting from inside with the series expansion in (\ref{eq:22SEM0}) slightly away from the origin. The characteristic transition region can be seen at some radius. A unique $p_\infty$ is found by requiring the numerical solution to approach an asymptotically flat behavior. 
The would-be horizon size $r_H$ is further obtained from a fit of the solution with a Schwarzschild metric at large $r$. \footnote{We do the fit at a large enough $r$, where the exponentially small corrections from $m_2$ are negligible.}

To have a better idea about the solution space, we can rewrite the metric as functions of a dimensionless ratio $r/R_0$, where $R_0$ denotes some typical size. As we will see below, a convenient choice of $R_0$ is the size of the 2-2-hole interior, which is $r_H$ or $r_a$ in different regions of the parameter space. The two field equations then become 
\begin{eqnarray}\label{eq:EoM2scale}
\frac{1}{R_0^2 \lp^2}[...]=0,\quad
\frac{1}{R_0^2 \lp^2}\left([...]+\frac{\lambda_2^2}{R_0^2}[...]+p_\infty R_0^2 \lp^2[...]\right)=0\,,
\end{eqnarray}
where $[...]$ denotes dimensionless quantities. This shows that a class of 2-2-hole solutions for $A, B$ as functions of $r/R_0$ is defined by $R_0/\lambda_2=\textrm{const.}$ and $p_\infty R_0^2 \lp^2=\textrm{const.}$. 
Since the two constants are related by the asymptotic flatness condition, the relativistic thermal gas model is described by a one-parameter family of solutions.  

\begin{figure}[!h]
  \centering%
{ \includegraphics[width=7.9cm]{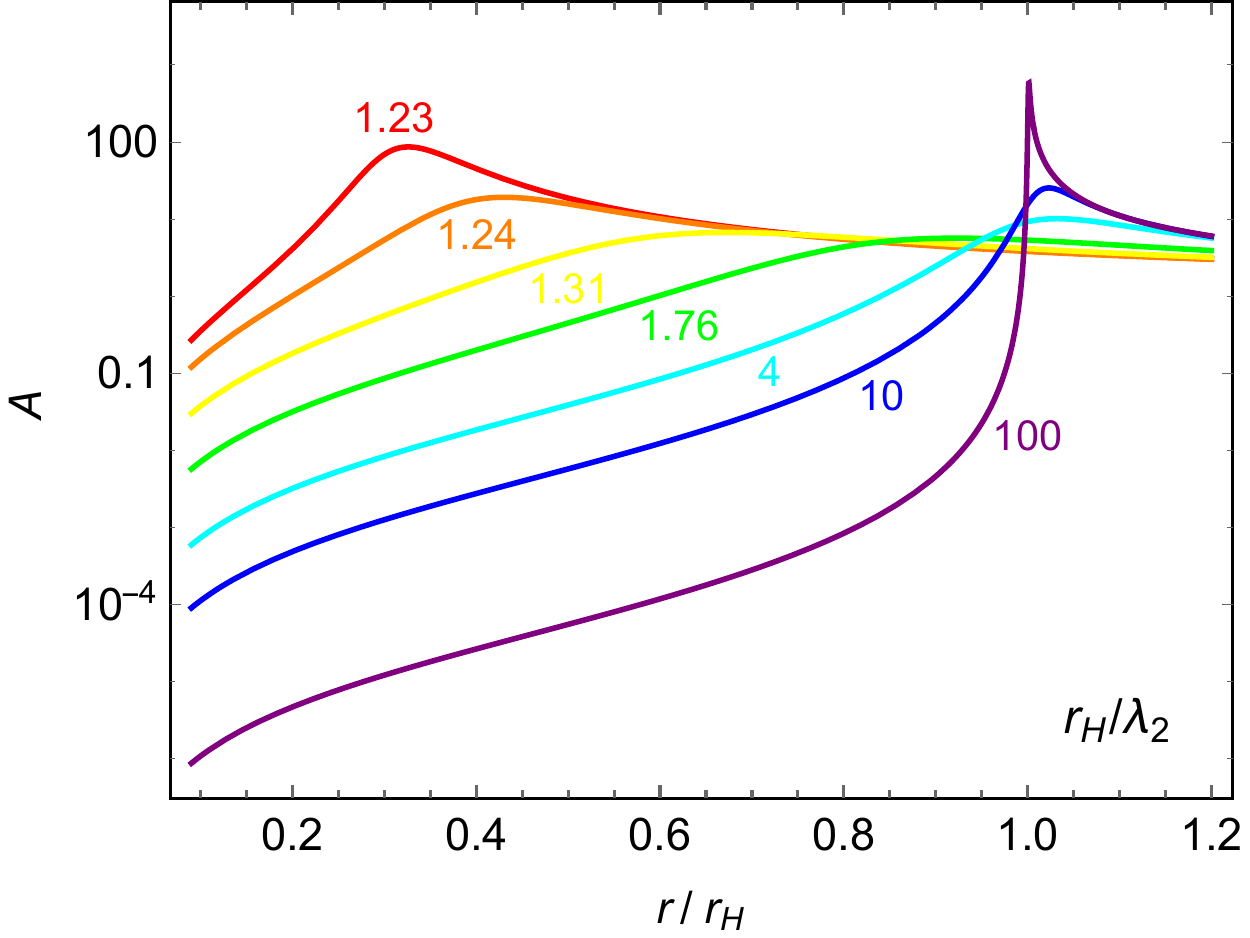}}\,\,\,
{ \includegraphics[width=7.9cm]{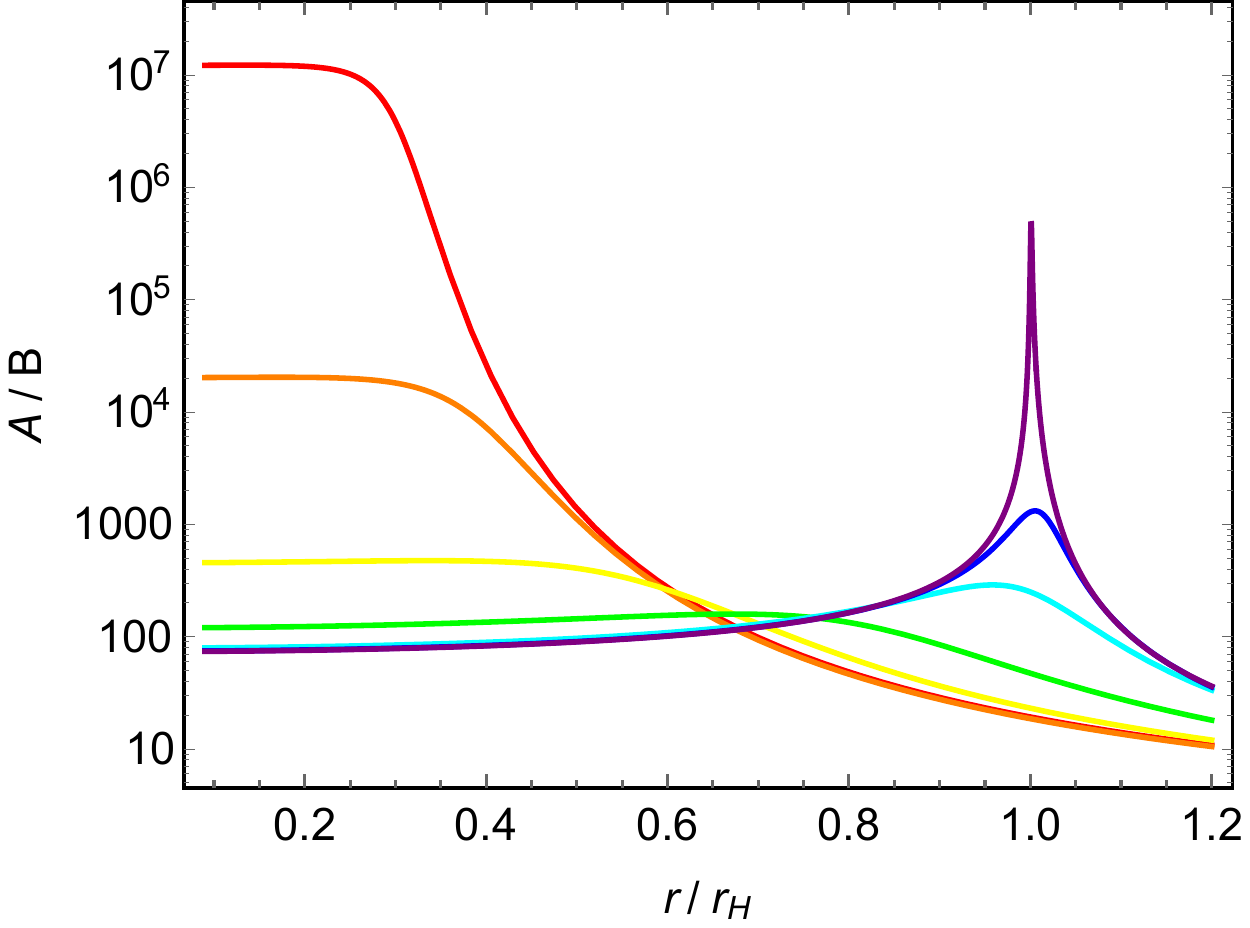}}\\
{ \includegraphics[width=7.9cm]{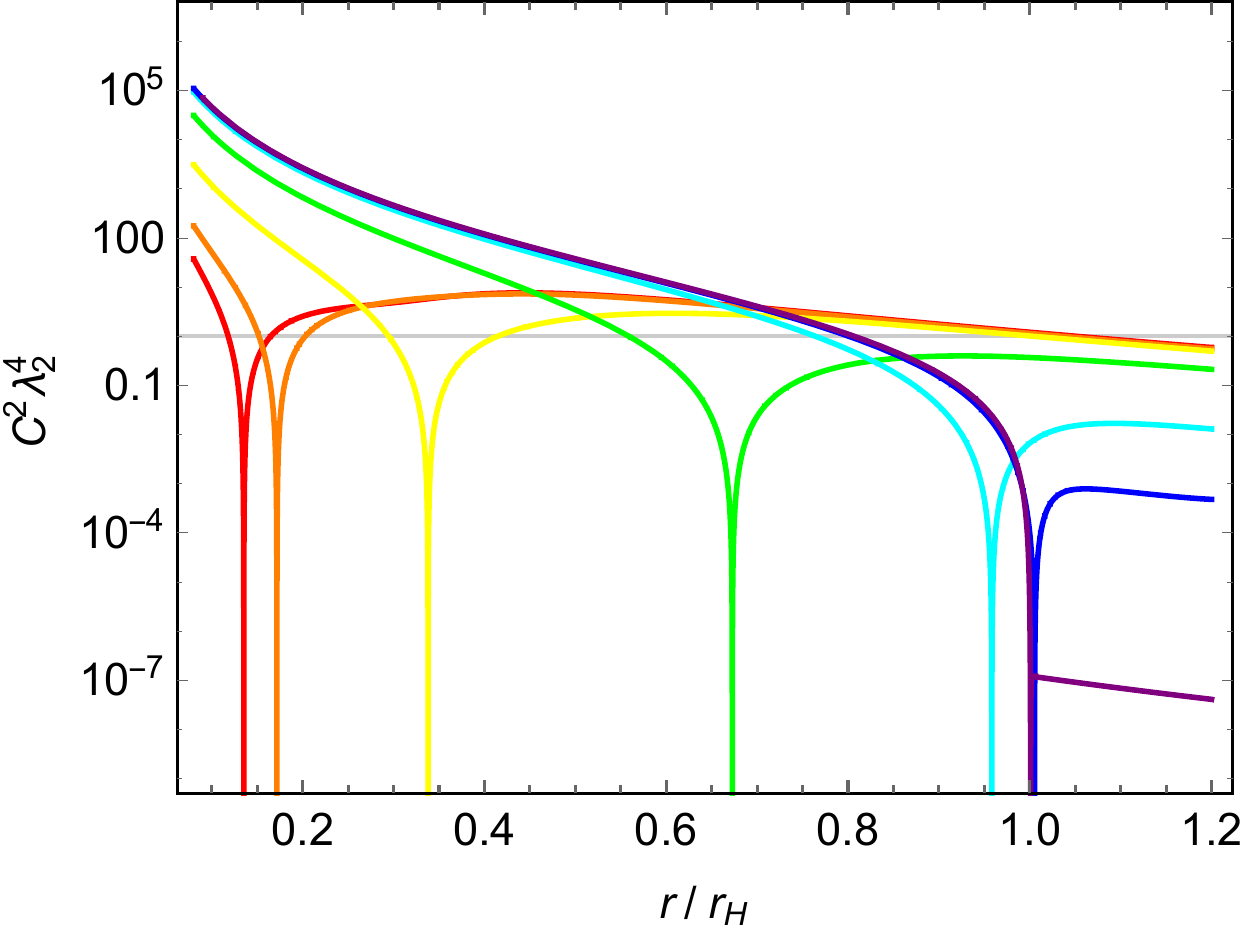}}\,\,\,
{ \includegraphics[width=7.9cm]{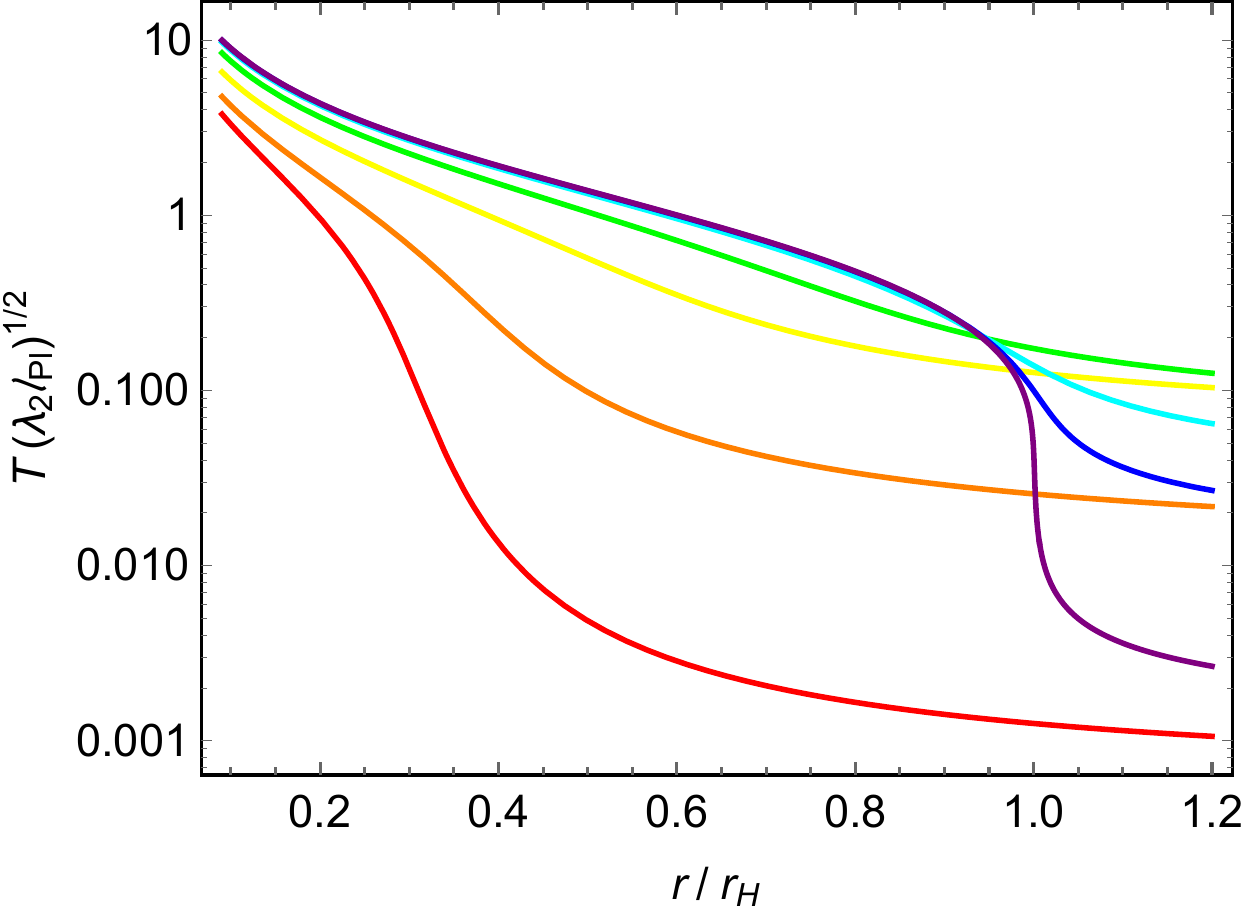}}
\caption{\label{fig:m2rH1} 
The metric $A$, the ratio $A/B$, Weyl tensor square $C^{\mu\nu\rho\sigma}C_{\mu\nu\rho\sigma}$ and the relativistic thermal gas temperature $T$ (for $N=1$) as functions of $r/r_H$ for $r_H/\lambda_2=$1.23 (red), 1.24 (orange), 1.31 (yellow), 1.76 (green), 4 (cyan), 10 (blue), 100 (purple).}
\end{figure}

Fig.~\ref{fig:m2rH1} presents metric and matter properties as functions of $r/r_H$ for selected solutions in the one-parameter family, as labelled by the value of $r_H/\lambda_2$. There is also a one to one correspondence to the value of $r_a/\lambda_2$ that increases with respect to $r_H/\lambda_2$. In the large mass limit, $r_H/\lambda_2\gg 1$ and $r_a/\lambda_2\gg1$, the solution has a narrow transition region around the would-be horizon, as characterized by the $A/B$ peak 
sitting slightly outside $r_H$. The ratio $\sqrt{B/A}$ denotes the radial speed of light and determines the time delay for external probes of the 2-2-hole interior~\cite{Holdom:2016nek}.  
The peak position is also where the Weyl tensor square $C^{\mu\nu\rho\sigma}C_{\mu\nu\rho\sigma}$ vanishes, given that $R=0$ for the Einstein-Weyl theory with $T_\mu^\mu=0$. The transition region connects the low curvature exterior as closely resembled by the Schwarzschild metric with the high curvature interior where $A/B$ quickly approaches a constant.  
With the object's size $r_H$ more comparable to the Compton wavelength $\lambda_2$, there are larger corrections around $r_H$, and $r_a$ drops rapidly. 
In the small mass limit, $r_H/\lambda_2\gtrsim 1$ and $r_a/\lambda_2 \ll1$, the $A/B$ peak is pushed well within the would-be horizon, and we see a broader transition region stretching roughly from $r_H$ to $r_a$. The interior shrinks drastically in all directions and the radial speed of light falls even faster. No solutions are found for $r_H/\lambda_2\lesssim  1$. Therefore, unlike many other ultracompact objects, a 2-2-hole can be arbitrarily heavy, but it has a minimum mass with $r_H/\lambda_2\sim 1$ and $r_a/\lambda_2\sim0$. 

For the thermal gas, its local temperature grows at smaller radius and reaches infinity at the origin. This corresponds to a quite compact matter distribution and naturally guarantees a 2-2-hole solution.  
If to compare with the thin-shell model~\cite{Holdom:2016nek}, it is similar to the case where the shell is  deep inside. 
Although the temperature blows up at the origin, the integrand in (\ref{eq:SUm0}) approaches a finite constant with $A, B\sim r^2$. The total entropy and energy for 2-2-holes, which shall be defined from the origin up to somewhere close to the would-be horizon, then remain finite. 
As mentioned before, $T_\infty$ can be viewed as the temperature measured at infinity no matter whether the gas remains in equilibrium or not outside the object. When $T_\infty$ is higher than that of the cosmic microwave background, the 2-2-hole will radiate like a normal object. 
For decreasing $r_H/\lambda_2$, we see $T_\infty$ first increases and then decreases, in contrast to a monotonic behavior for a black hole.

Apparently 2-2-holes in the large and small mass limits are qualitatively different. Moreover they are governed by distinct scaling behaviors at the leading order, which relate the interior solutions at different $r_H/\lambda_2$. Tab.~\ref{tab:scaling} summarizes some essential properties of spacetime and matter under two different scalings. 
Basically four (five) length scales are relevant. In the large mass limit, the 2-2-hole is characterized by the would-be horizon size $r_H$ and the Compton wavelength $\lambda_2$. In the small mass limit, the behavior is controlled by the new scale $r_a$ instead (and the $B(r)$ normalization scale $r_b\equiv 1/\sqrt{b_2}$).
For both cases, matter properties depend on the Planck length $\lp$ in addition. 
For comparison, we also include the naive scaling for self-gravitating radiation in a box of radius of order $r_H$. 
In the following, we discuss the features in Fig.~\ref{fig:m2rH1}, the meaning of Tab.~\ref{tab:scaling} and their physical implications in detail for these two limits respectively.  

\begin{table}[h]
\centering 
\caption{Novel scaling behaviors for the 2-2-hole interior in the large and small mass limits}
\vspace{0.5em}
\renewcommand{\arraystretch}{1.5}
\begin{tabular}{c | c | c c c | c c c}
\hline\hline
& Scales relation & Metric & Curvature & $\ell_\textrm{in}$ & Temperature 
& $U_\textrm{in}$
& $S_\textrm{in}$
\\
\hline
\multirow{2}{*}{Large mass} & 
\multirow{2}{*}{$r_a\gg r_H\gg \lambda_2$} & 
\multirow{2}{*}{$\displaystyle{A(r)\frac{r_H^2}{\lambda_2^2}}$, $\displaystyle{B(r)\frac{r_H^2}{\lambda_2^2}}$} & 
\multirow{2}{*}{$I(r)\lambda_2^{d_I}$} & 
\multirow{2}{*}{$\lambda_2$} & 
\multirow{2}{*}{$\displaystyle{T(r)\sqrt{\lambda_2\lp}}$} & 
\multirow{2}{*}{$\displaystyle{\frac{r_H}{\lp^2}}$}  & 
\multirow{2}{*}{$\displaystyle{\frac{r_H^2}{\lp^2}\sqrt{\frac{\lp}{\lambda_2}}}$} 
\\
&&&&&&&\\
\hline
\multirow{2}{*}{Small mass} & 
\multirow{2}{*}{$r_H\gtrsim \lambda_2 \gg r_a$} & 
\multirow{2}{*}{$A(r)$,  $\displaystyle{B(r)\frac{r_b^2}{r_a^2}}$} & 
\multirow{2}{*}{$I(r)r_a^{d_I}$} & 
\multirow{2}{*}{$r_a$} & 
\multirow{2}{*}{$T(r)\sqrt{r_a \lp}$} & 
\multirow{2}{*}{$\displaystyle{\frac{1}{r_b}\frac{r_a^2}{\lp^2}}$} & 
\multirow{2}{*}{$\displaystyle{\left(\frac{r_a^2}{\lp^2}\right)^{3/4}}$}
\\
&&&&&&&\\
\hline\hline
\multirow{2}{*}{Naive scaling} & 
\multirow{2}{*}{/} & 
\multirow{2}{*}{$A(r)$,  $B(r)$} & 
\multirow{2}{*}{$I(r)r_H^{d_I}$} & 
\multirow{2}{*}{$r_H$} & 
\multirow{2}{*}{$T(r)\sqrt{r_H \lp}$} & 
\multirow{2}{*}{$\displaystyle{\frac{r_H}{\lp^2}}$}  & 
\multirow{2}{*}{$\displaystyle{\left(\frac{r_H^2}{\lp^2}\right)^{3/4}}$}
\\
&&&&&&&\\
\hline\hline
\end{tabular}
\label{tab:scaling}
\end{table}

\subsection{Large mass limit}
\label{sec:largeM}

The novel scaling behavior for large 2-2-holes with $r_H/\lambda_2\gg1$ has already been noticed before~\cite{Holdom:2016nek}. Here we generalize previous results by keeping the full dependence on all dimensional scales, without assuming $m_2\sim\Mp$. 
This limit also applies to a given size 2-2-hole when $\alpha\to 0$, i.e. the decoupling limit of the massive mode. \footnote{In the quantum theory, $\lambda_2\ll \lp$ implies large quantum corrections from strong gravitational couplings, and the decoupling limit in the classical theory might be irrelevant.}
The typical size $R_0$ is identified with $r_H$ for the 2-2-hole interior. At the leading order, the following dimensionless quantities are found to be only functions of $r/r_H$,
\begin{eqnarray}\label{eq:LMscaling1}
A(r) \frac{r_H^2}{\lambda_2^2},\quad
B(r) \frac{r_H^2}{\lambda_2^2},\quad
I(r)\lambda_2^{d_l}\,,
\end{eqnarray}
where $I(r)$ denotes some curvature invariant of dimension $d_I$. That is, for different 2-2-holes in the one-parameter family with $r_H/\lambda_2\gg1$, these combinations as functions of $r/r_H$ appear the same for the interior to a very good approximation.
This corresponds to $a_2, b_2\propto \lambda_2^{2}/r_H^{4}$ for the expansion coefficients, and the scales relation $r_a\gg r_H\gg \lambda_2$. 
So the metric $A(r)$ falls rapidly in the large mass limit, and the radial proper length for the interior scaling region $\ell_\textrm{in}$ reduces dramatically in comparison to the angular one $\sim r_H$. $\ell_\textrm{in}$ actually becomes decoupled from $r_H$ at the leading order and is only of the order of $\lambda_2$. 
The curvature invariants at the interior boundary $I(r_H)$ are fixed by $\lambda_2$ as well. 
These novel properties of interior geometry result from the interplay between its size $r_H$ and the external scale $\lambda_2$.
The scaling of $B(r)$ is determined by the normalization rather than field equations, and it turns out to be the same as $A(r)$. 
The radial speed of light in the interior then doesn't change for different $r_H/\lambda_2$. 

With increasing $r_H/\lambda_2$, the boundary of the interior scaling region moves towards $r_H$  and the transition region becomes narrower. The later is characterized by the $A/B$ peak at $r_\textrm{peak}\equiv r_H(1+\delta)$ (also roughly the place that a significant deviation from Schwarzschild metric can be seen) that is slightly outside the would-be horizon, with $\left(A/B\right)_\textrm{peak}\sim \delta^{-2}$. From numerical solutions we find a rapidly growing peak with $1/\delta \sim (r_H/\lambda_2)^{\eta}$ and $\eta\in [1.75, 2]$~\cite{Holdom:2016nek}. This implies a dip of the radial speed of light, and the transition region gives the dominant contribution to the time delay for an external probe. 
Fortunately the time delay only has a logarithmic dependence on the large ratio, $r_H \log (r_H/\lambda_2)$.
For an astrophysically large 2-2-hole, it remains quite accessible, i.e. $M_\odot \log (M_\odot/\Mp)\sim \mathcal{O}(1\textrm{ms})$. 
There is then the hope to detect a Planckian distance deviation outside a macroscopic horizon by gravitational wave echoes.

For the relativistic thermal gas, its interior local temperature $T(r)\sqrt{\lambda_2 \lp}$ is a function of $r/r_H$.
With metric properties in (\ref{eq:LMscaling1}), we find the following scalings for the temperature at infinity and  the interior contribution to the total entropy and energy,
\begin{eqnarray}\label{eq:TSLmass}
T_\infty \propto \frac{1}{r_H}\sqrt{\frac{\lambda_2}{\lp}},\quad
S_\textrm{in}\propto \frac{r_H^2}{\lp^2}\sqrt{\frac{\lp}{\lambda_2}},\quad
U_\textrm{in}
\propto \frac{r_H}{\lp^2}\,.
\end{eqnarray}
Some familiar properties of black holes thermodynamics, $T_\infty\propto 1/r_H$, $S\propto r_H^2$, $U\propto M$, now arise as a result of classical thermodynamics of ordinary matter source for the 2-2-hole background.
To see the difference more clearly, it is useful to make comparison with the naive expectation.
Imagine a black hole formed by compressing a relativistic thermal gas into a volume $V\propto r_H^3$ as defined by the horizon size. With the naive scaling, the gas contribution to the black hole energy and entropy has: $U_\textrm{in}\propto V T^4$ and $S_\textrm{in}\propto V T^3$. To have $U_\textrm{in}\propto M$, we find $T\propto (r_H\lp)^{-1/2}$ and then $S_\textrm{in}\propto (r_H/\lp)^{3/2}$~\cite{Sorkin:1981wd}.
When $r_H\gg \lp$, this is apparently too small to account for the Bekenstein-Hawking entropy for the black hole. 
While for the 2-2-hole, the novel scaling leads to quite different expressions: $U_\textrm{in}\propto r_H \lambda_2^2 T^4$, $S_\textrm{in}\propto r_H^2 \lambda_2 T^3$, and the enormous entropy scaled with the area and $U_\textrm{in}\propto M$ can be realized simultaneously as in (\ref{eq:TSLmass}). 
Two factors play the role here.
Firstly, large gravitational redshift in the interior gives additional suppression to energy in comparison to entropy. As a result, the hot gas temperature doesn't decrease with the object's large size $r_H$. 
Secondly, with $\ell_\textrm{in}\propto \lambda_2$, the squeezed interior volume receives the dominant contribution from the region close to the boundary and scales effectively as the area.
Therefore, we see an intriguing connection to the original attempts that attribute the black hole entropy to its thermal atmosphere around the horizon. In contrast to the old proposals, such as the brick wall model~\cite{tHooft:1984kcu} and the stretched horizon~\cite{Price:1986yy}, the thermal gas responsible for entropy here is exactly the matter source for the background spacetime, with the distribution fully determined by the conservation law. Many problems are then avoided.

With explicit dependence on the ratio $\lp/\lambda_2$ and the number of particle species $N$, the  numerical values of thermodynamic variables are different from black holes in general. The 2-2-hole entropy and energy can be defined by the integration in (\ref{eq:SUm0}) up to $r\sim\mathcal{O}(r_H)$, and it is still dominated by the interior contribution. Including numerical factors, we find
\begin{eqnarray}\label{eq:LMentropy}
T_\infty\approx 1.35\, N^{-1/4}\sqrt{\frac{\lambda_2}{\lp}}T_H,\quad
S\approx 0.74\, N^{1/4}\sqrt{\frac{\lp}{\lambda_2}}S_\textrm{BH},\quad
U=\frac{3}{4}S T_\infty \approx \frac{3}{8}M\,.
\end{eqnarray}
$T_H=1/4\pi r_H$ is the Hawking temperature and $S_\textrm{BH}=\pi r_H^2/\lp^2$ is the Bekenstein-Hawking entropy.  Although $S, T_\infty$ depend on $N$, $\lambda_2$ explicitly, their product $S T_\infty\propto p_\infty$, which sources the background spacetime, does not. The scaling law in the large mass limit further indicates $S T_\infty \propto M$. Interestingly, the numerical value is found to approach that for the black hole with precision,  
\begin{eqnarray}\label{eq:STproduct}
S T_\infty = S_\textrm{BH}T_{H}=\frac{M}{2}\,.
\end{eqnarray}
This leads to the same $U$ as in the brick wall model, where the UV cutoff is chosen to reproduce the Bekenstein-Hawking entropy~\cite{tHooft:1984kcu}.
Therefore, the physical mass $M$ of a large 2-2-hole has a sizable fraction from the gas energy $U$, with the gravitational field contribution being comparable. 
The numerical coincidence (\ref{eq:STproduct}) is also related to the first law of thermodynamics,
\begin{eqnarray}\label{eq:firstlaw}
\frac{d S}{S}=2\frac{d r_H}{r_H}=2\frac{d M}{M} \Rightarrow 
d M=\frac{M}{2S}dS=T_\infty dS\,.
\end{eqnarray} 
This is expected given that $S$ is no different from a normal entropy.
\footnote{In principle there shall be the work term $\sqrt{B}\,p \,dV$ from the gas in (\ref{eq:firstlaw}). But with the boundary $\sim\mathcal{O}(r_H)$ in the exterior region, $\sqrt{B}\,p \,dV \propto (\lambda_2^2/r_H^2)\, dM$ is highly suppressed when $r_H/\lambda_2\gg1$ and is negligible.}  
Similarly a configuration with larger entropy shall be more stable from the second law of thermodynamics. 
In the strong coupling scenario $m_2\approx \Mp$, the entropy of a 2-2-hole can easily surpass that of the same mass black hole with a reasonable $N$, say for the Standard Model particles. So a 2-2-hole 
rather than a black hole, which is still a solution in classical quadratic gravity, is more likely to be the endpoint of gravitational collapse.\footnote{Here we restrict to 2-2-holes with only even power terms in the series expansion. More general solutions could have a larger entropy, but still with similar features.}

Given the absence of a horizon, we shall also expect the generalized second law of thermodynamics for the 2-2-hole.
It is nonetheless interesting to see how the general argument used for black holes works here~\cite{Wald:1995yp, Wald:1999vt}. Imagine lowering a box of matter with entropy $S_m$
 toward the ultracompact object and allowing matter fall into the object at some point. A distant observer has to hold the box (say with a rope) to make this an quasi-static process. Assuming no generation of entropy, the optimal place to release the matter is to have the minimal energy deposit to the object and then the minimal entropy increase. Considering also the pressure from the thermal gas, this place is found somewhere outside the would-be horizon, with the minimal mass increase $(\Delta M)_\textrm{min}=T_\infty V s$, where $s$ is the local entropy density for the thermal gas and $V$ is the box volume. 
With the first law (\ref{eq:firstlaw}), the minimal entropy increase $(\Delta S)_\textrm{min} =(\Delta M)_\textrm{min}/T_\infty = V s$ remains the same as that for the black hole~\cite{Wald:1995yp}. Given that the relativistic thermal gas has the maximum entropy at a given energy and volume, the minimal change of total entropy $-S_m + V s$ stays positive.

So far all discussions of thermodynamics are for the classical source of the 2-2-hole. Once including quantum corrections, the renormalized vacuum energy density normally gives additional contribution to the stress tensor. For a horizonless and static spacetime, the Boulware vacuum is a natural choice. It is defined with respect to the Killing time, and has growing negative vacuum energy density in the high redshift region. To have negligible backreaction to the metric, a ``topped-up" Boulware state with hot quantum fields excitations is constructed, where the field temperature is roughly $T_H$~\cite{Mukohyama:1998rf}.
The thermodynamics of hot quantum fields is very similar to what we derived here for the thermal gas~\cite{Holdom:2016nek}. 
With this additional contribution, $S, U$ shall be larger than what we found above for the classical gas source.

\subsection{Small mass limit}
\label{sec:smallM}

When $r_H$ becomes comparable to $\lambda_2$, the metric $A(r)$ grows large in the interior and we see significant deviation from the large mass behavior in Fig.~\ref{fig:m2rH1}. In this small mass limit, the 2-2-hole interior size $R_0$ is characterized by the decreasing scale $r_a$ instead.
As we can see from the series expansion in (\ref{eq:22SEM0}), contribution from $\lambda_2$ becomes negligible when $r_a/\lambda_2\ll1$. So a different scaling law emerges, with $r_a$ the only essential scale. 
At the leading order, we find following dimensionless quantities being functions of $r/r_a$,\footnote{The Weyl tensor square is one exception. With a cancellation at the leading order, $C^2\sim 27m_2^4r_a^4/r^4$ is a function of $r/r_a$.}
\begin{eqnarray}\label{eq:SMscaling1}
A(r),\quad
B(r)\frac{r_b^2}{r_a^2},\quad
I(r) r_a^{d_l}\,.
\end{eqnarray}
The radial proper length for the interior $\ell_\textrm{in}\propto r_a$ now scales the same as the angular one. The curvature invariant at the interior boundary $I(r_a)$ is characterized by $r_a$ as well. Except for the $B(r)$ normalization as defined by $r_b\equiv 1/\sqrt{b_2}$, (\ref{eq:SMscaling1}) is quite similar to the naive scaling that is governed by a unique length scale, and it can be deduced from (\ref{eq:LMscaling1}) with both $r_H, \lambda_2$ identified as $r_a$.
The normalization scale $r_b$ is found to increase exponentially for smaller $r_a$, with $r_b/r_a\propto \exp\left(\mathcal{O}(1)\lambda_2^2/r_a^{2}\right)$. This implies an extremely deep gravitational potential in the small 2-2-hole interior and a tremendous redshift. 

We find that the would-be horizon $r_H$ is quite insensitive to $r_a$ in the small mass limit. 
As less compact objects, visible deviations from the Schwarzschild metric occur at a distance of the order of $r_H$ outside the would-be horizon. 
With the $A/B$ peak pushed further from $r_H$ and more towards $r_a$, the broad transition region stretches roughly from $r_H$ to $r_a$, and the curvature is characterized by $\lambda_2\sim r_H$. 
As shown in Fig.~\ref{fig:m2rH1}, the transition region has a quite asymmetric shape.
The ratio $A/B$ still quickly approaches a constant in the interior, but its value $\sim r_b^2/r_a^2$  now grows together with the large peak due to different scalings of $A, B$. This corresponds to an exponentially falling radial speed of light. A significant amount of time delay for an external probe then comes from the interior. Since it scales exponentially with the large ratio $\lambda_2/r_a$, as opposed to the large mass limit, the time delay will soon surpass the age of the universe.  

The interior temperature for relativistic thermal gas remains high, with the combination $T(r)\sqrt{r_a \lp}$ a function of $r/r_a$. The scalings for the temperature at infinity and the interior contribution to entropy and energy now go like,
\begin{eqnarray}\label{eq:TSSmass}
T_\infty\propto \frac{1}{r_b}\sqrt{\frac{r_a}{\lp}},\quad
S_\textrm{in}\propto \left(\frac{r_a^2}{\lp^2}\right)^{3/4},\quad
U_\textrm{in}\propto \frac{1}{r_b}\frac{r_a^2}{\lp^2}.
\end{eqnarray}
Due to the large redshift, $T_\infty$ drops dramatically for smaller objects. This leads to a positive heat capacity, $dM/dT_\infty>0$, as classical thermodynamic systems. 
For the entropy, the interior contribution still dominates over the transition region. Since the radial and angular proper lengths both scale with $r_a$, the area law no longer applies. Actually, the entropy appears just like what we naively expect and falls rapidly with the size of small objects. 
Given the discrete nature of particles, $S$ might not approach zero continuously. The minimum 2-2-hole with nonvanishing entropy shall have $S\approx 1$. 
The product $S T_\infty$ now almost decouples from $r_H$ and becomes far smaller than the physical mass $M$. 
In other words, the gravitational field energy $\sim M$ dominates in this limit and the gas contribution $U\propto S T_\infty$ is almost negligible.
The first law of thermodynamics $dM\approx T_\infty dS$ is still approximately true.
\footnote{The work term $\sqrt{B}\,p\, dV \propto (r_H^2r_a^2/r_b^4)\, dM$ is even more suppressed in the small mass limit and totally negligible.}  
 Given the exponentially falling $T_\infty$ and numerical errors in fitting $r_H$, the first law can only be confirmed at a much worse precision. But assuming its validity, $d r_H\propto (r_a/r_b)\,d r_a$, we can  find $r_H$ as a function of $r_a$. When $r_a/\lambda_2$ approaches zero, $r_H$ does vary extremely slow with  $(r_H-r_{H,\,\textrm{min}})/\lambda_2\propto (r_a/\lambda_2)^{3}\exp\left(-\mathcal{O}(1)\lambda_2^2/r_a^{2}\right)$. 

A large 2-2-hole hotter than the cosmic microwave background will radiate like a black hole. The evaporation becomes faster for smaller objects due to the negative heat capacity. The behavior changes at $r_H/\lambda_2\approx 1.5$ and $r_a/\lambda_2\approx 1.4$, when $T_\infty$ reaches a maximum around $0.1\, N^{-1/4}\sqrt{m_2\Mp}$. Below this temperature, the heat capacity turns positive, and both the evaporation rate and the temperature drop significantly. So instead of an explosion in the standard picture for black holes, a small 2-2-hole becomes colder and evaporate slower at the late time. 
At certain point, it appears stable for the age of the universe. The cold remnant with mass $M_\textrm{min} \sim \Mp^2/m_2$ could then serve as a dark matter candidate. 
In the strong coupling scenario of quantum theory, $m_2\sim \Mp$ and $M_\textrm{min}\sim\Mp$, this naturally gives rise to a Planck-sized remnant. While in the weak coupling case with $m_2\ll \Mp$, the remnant could be much larger and heavier. Different thermodynamic behaviors for small 2-2-holes may have interesting implications for dark matter phenomenology.
~\footnote{Primordial black hole mimickers as dark matter candidates are also discussed for nonlocal stars~\cite{Buoninfante:2019swn} and wormholes~\cite{Damour:2007ap, Berthiere:2017tms}.}   

The existence of a lower mass bound for 2-2-holes may have some relation to the uncertainty principle. As a familiar candidate for exotic compact objects, a boson star is commonly viewed as a macroscopic quantum state controlled by the uncertainty principle~\cite{revBS1, revBS2}. The simplest case is a mini-boson star composed of free massive scalar bosons, which reaches the minimal radius at the maximal mass (the stable branch) with $R_\textrm{min}\sim M_\textrm{max}/\Mp^2$. The fact that the object size is no smaller than the Compton wavelength of the massive modes then implies $M_\textrm{max}\sim \Mp^2/m_0$. A 2-2-hole reaches the minimal radius at the minimal mass instead. A similar form of mass $M_\textrm{min}\sim \Mp^2/m_2$ then suggests a direct connection between the lower bound $r_H/\lambda_2\gtrsim1$ and the uncertainty principle.

\section{Variations}
\label{sec:variation}

\subsection{General thermal gas model}
\label{sec:thegas2}

The most straightforward generalization of the relativistic thermal gas model with $\rho=3p$ is to include nonzero particle mass. The energy density and pressure in (\ref{eq:stresstensor}) are then 
\begin{eqnarray}\label{eq:rhop2}
\rho=\frac{3N T^4}{\pi^2}f_\rho\left(\frac{m}{T}\right),\quad
p=\frac{N T^4}{\pi^2}f_p\left(\frac{m}{T}\right).
\end{eqnarray}
As before we approximate $f_\rho(x), f_p(x)$ by expressions for $\epsilon=0$ case,
\begin{eqnarray}
f_\rho(x)\approx\frac{1}{6}x^2(x K_1(x)+3K_2(x)),\quad
f_p(x)\approx\frac{1}{2}x^2K_2(x)\,,
\end{eqnarray}
with small difference for $\epsilon=\pm1$ absorbed in $N$. $K_\alpha(x)$ is modified Bessel function of the second type. In the zero mass or infinite temperature limit, $f_\rho(0)=f_p(0)=1$ and $\rho=3p$ is recovered.

Since the ratio $\rho/p$ is $r$ dependent for nonzero mass, three variables $A(r), B(r), T(r)$ need to be simultaneously solved from field equations (\ref{eq:EoM1}) and the conservation law (\ref{eq:consL1}),
with $T_\mu^\mu=3p-\rho$ and $\rho, p$ defined by $T$ in (\ref{eq:rhop2}).
As long as the temperature blows up at the origin, the leading order behavior of the series expansion remains the same as the massless case,  
\begin{eqnarray}
A(r)=a_2r^2+...,\quad
B(r)=b_2r^2+...,\quad
T(r)=\frac{c_{-1}}{r}+...\,,
\end{eqnarray}
and mass corrections only enter at subleading orders. For a given $m$ and  $(\lambda_2, r_a)$, we find the numerical solution in a similar way by shooting from inside, and $c_{-1}$ is fixed by the demanding asymptotic behavior. If all variables are written as functions of $r/R_0$, field equations would have similar structure as (\ref{eq:EoM2scale}), with $p_\infty$ replaced by $T^4f_i(m/T)$. A class of solutions for $A, B, T(R_0\lp)^{1/2}$ as functions of $r/R_0$ is then defined by three dimensionless constants $R_0/\lambda_2, m\,(R_0 \lp)^{1/2}$ and $c_{-1}(\lp/R_0)^{1/2}$. With the asymptotic flatness condition, the 2-2-hole sourced by a general thermal gas is a two parameters family of solutions.

\begin{figure}[!h]
  \centering%
{ \includegraphics[width=8.05cm]{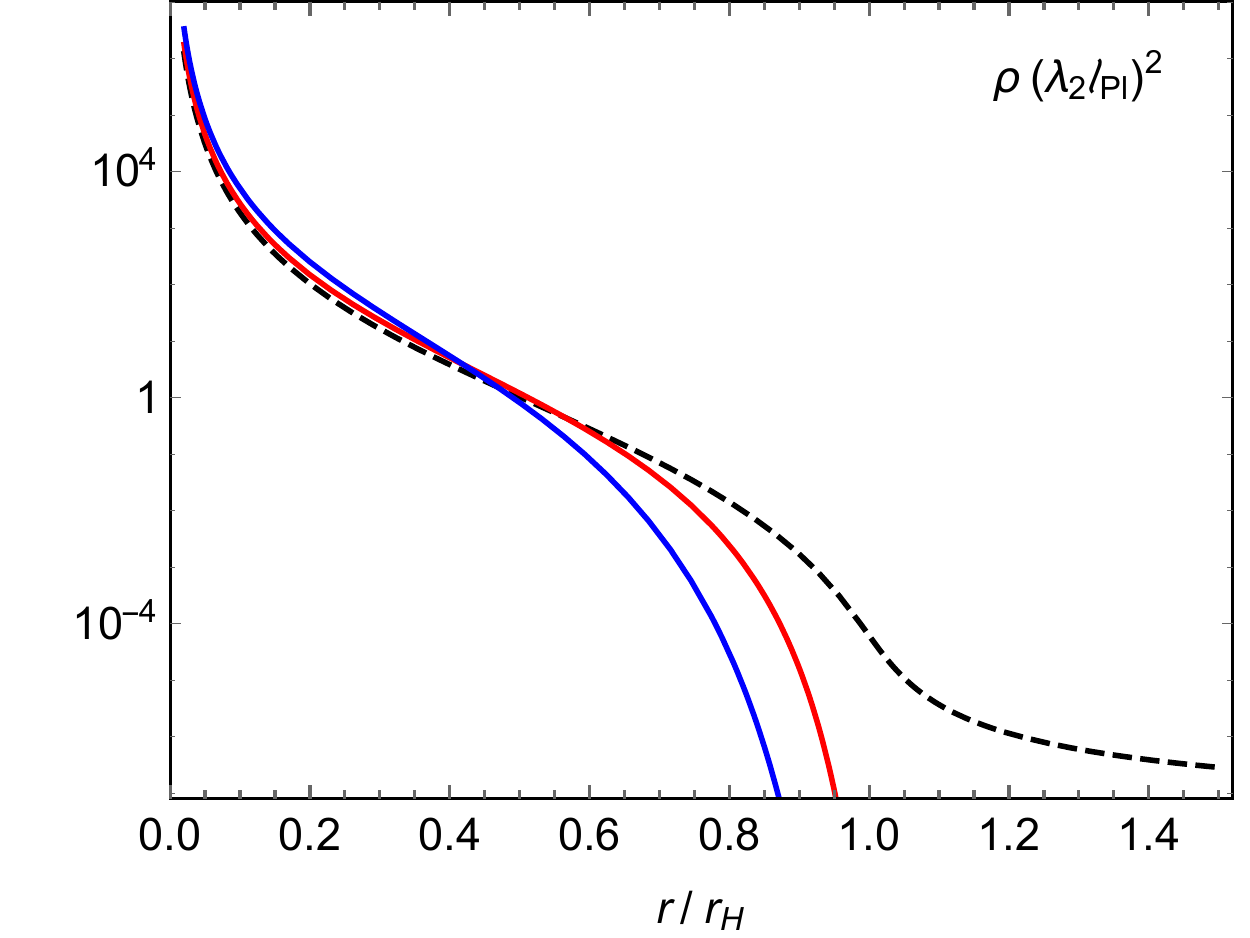}}\,\,\,\,\,
{ \includegraphics[width=7.8cm]{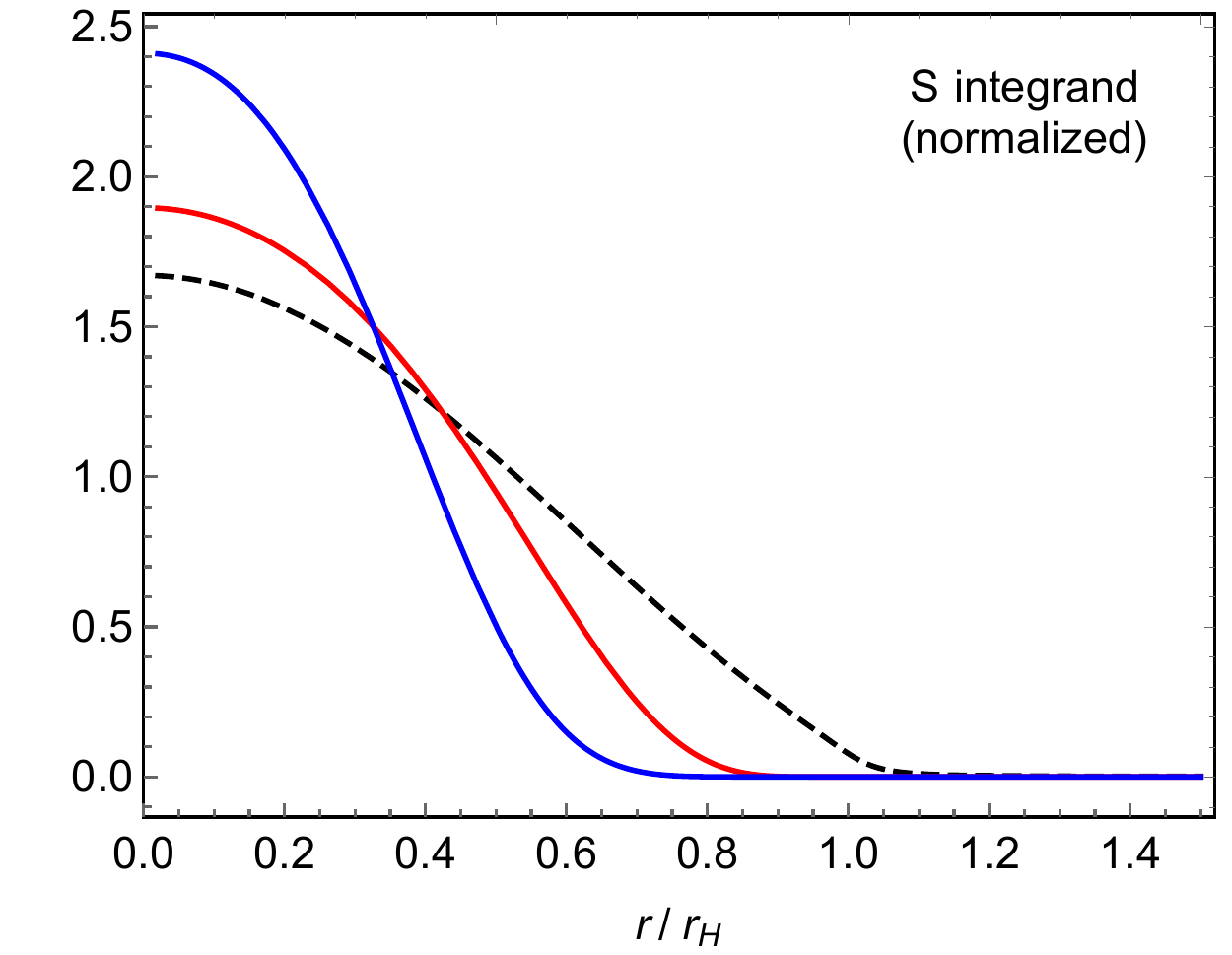}}
\caption{\label{fig:mass} 
The energy density and the entropy integrand normalized by $S$ for the massless case as function of $r/r_H$, for $r_H/\lambda_2=6$ and $m\,(r_H\lp)^{1/2}=0$ (black dash), $7$ (red), $20$ (blue).}
\end{figure}

Nonzero mass corrections become important when $m$ is comparable to the gas temperature at some radius in the interior. Outside this radius,  the gas becomes non-relativistic and the ratio $p/\rho$ drops significantly. To reach the same physical mass, a higher interior temperature (a larger $c_{-1}$) is needed, and a larger density in the relativistic region compensates the declining contribution from the non-relativistic region, as in Fig.~\ref{fig:mass}. 
We note that the relation $T(r)B(r)^{1/2}=\textrm{const.}$ remains a very good approximation for a general $m$. This can be seen by solving the conservation law in the opposite limit $m\gg T$. At the leading order, with $\rho/p=m/T$ and $p\propto T^4(m/T)^{3/2}\exp(-m/T)$, (\ref{eq:consL1}) is reduced to $T'/T+B'/2B=0$ too. The Tolman's law then applies at both small and large radii. 
Since the mass dependence comes mainly through the ratio $m/T$ in the interior, there are the same scaling behaviors as in Tab.~\ref{tab:scaling} for a given $m\,(\lambda_2\lp)^{1/2}$ and $m\,(r_a\lp)^{1/2}$ in the large and small mass limits respectively.  

We can calculate the total entropy $S$ for the general thermal gas model with the same formula (\ref{eq:entropy}). There is still the area law in the large mass limit $r_H/\lambda_2\gg1$. As shown in Fig.~\ref{fig:mass},  with increasing $m$, the contribution from the relativistic region goes up and the one from the non-relativistic region drops down. The total entropy results from their competition. 
Within numerical errors, we find that various thermodynamic variables remain quite close to their values for the massless case (\ref{eq:LMentropy}).\footnote{We have checked this for $r_H/\lambda_2\sim 5-100$ and $m\,(r_H \lp)^{1/2}\sim 0-8$. It is harder to find numerical solutions for a large $m$ because the shrinking relativistic region requires a smaller starting point and a higher accuracy for the shooting method.}
For example, $T_\infty/T_H$ , $U/M$ increase very slowly with $m$, while $S/S_\textrm{BH}$ decreases with a similar speed if to satisfy the first law of thermodynamics.
It is safe to say that thermodynamic variables for the 2-2-hole is quite insensitive to the particle mass, although the interior matter distributions do vary drastically. 
In reality a thermal gas may include particles of different masses. The Standard Model particles can be treated as massless if $(m_2\Mp)^{1/2}\gtrsim 1$\,TeV. New heavy particles that are inaccessible with the existing experimental techniques might be trapped in the 2-2-hole interior, but have little impact for the outside.  

As a final remark, the 2-2-holes can be sourced by more general matter distributions. 
For a spherically symmetric configuration, the stress tensor may naturally have anisotropic pressure $p_r \neq p_\theta$ as in (\ref{eq:stresstensorG}). As long as the stress tensor is traceless, we find the same leading order behavior for the series expansion as the thermal gas model, i.e. $\rho, p_r, p_\theta\propto 1/r^4$. A larger $p_\theta/p_r$ corresponds to a more compact matter distribution. When $p_\theta/p_r$ is not too small, we do find 2-2-hole solutions. Various parameters can change by some amount, but crucial features like the scaling behavior remain similar as for the thermal gas model. 
A more physical model of anisotropic pressure is a complex scalar theory with $\mathcal{L}_\phi=\frac{1}{2}\partial^\mu\phi^\dag\partial_\mu\phi+V(\phi)$~\cite{revBS2}. Solving the scalar field equation of motion under series expansion, the leading order behavior for the stress tensor can only be $1/r^6$ or $1/r^2$. The former is too singular to maintain the original $r^2$ expansion for the metric. While the later seems too soft and we fail to find numerical solutions by the shooting method. It seems the behavior $\rho, p_r, p_\theta\propto 1/r^4$ is crucial for the existence of 2-2-holes for continuous matter sources. More details on the anisotropic stress tensor can be found in Appendix.~\ref{sec:appC}.

\subsection{Perturbation from a matter shell}
\label{sec:shell}

Once a 2-2-hole has been formed, it can grow with matter accretion. It has been argued that the compactness of horizonless objects can be constrained in some rather model-independent way from its response to matter accretion~\cite{Carballo-Rubio:2018vin}. 
We want to see how a 2-2-hole will grow, and the relevance of these general arguments. 

For this purpose, we perturb a 2-2-hole sourced by the relativistic thermal gas by a matter shell at some radius $\ell$, to model a slow accretion of matter. Since the perturbed configuration receives the dominant contribution from the relativistic thermal gas,  we can find a series of static 2-2-hole solutions for a wide range of $\ell$ as long as the thin-shell mass remains subdominant. This is in contrast to our previous thin-shell model, where 2-2-holes only exist when $\ell\lesssim r_H$~\cite{Holdom:2016nek}.  

For simplicity, we consider a narrow width perturbation with vanishing radial pressure, i.e. $\Delta T_{\mu\nu}=\textrm{diag}\,(B\rho_\ell, 0, r^2 p_\ell, r^2 s_\theta^2 p_\ell)$, which satisfies the momentum conservation law: $p_\ell/\rho_\ell=r B'/4B$. The field equations (\ref{eq:EoM1}) then become, 
\begin{eqnarray}\label{eq:EoM3}
H_1=8\pi \frac{r B'-2B}{2B}\rho_\ell,\quad 
H_2=8\pi \frac{A }{B^2} p_\infty-8\pi\left(X B\rho_\ell+Y(B \rho_\ell)'\right)\,.
\end{eqnarray}
Here the perturbation is assumed to be a Gaussian density profile with some small width $\sigma$, $\rho_\ell(r)=\frac{d_\ell}{\sqrt{2\pi \sigma^2}} \exp\left[-\frac{1}{2}(r-\ell)^2/\sigma^2\right]$, which approaches a thin-shell when $\sigma\to0$. Assuming negligible interaction between the relativistic thermal gas and the shell, we keep $c_{-1}$ the same as the original unperturbed 2-2-hole with the would-be horizon $r_H$. For each $\ell$ and $d_\ell>0$, we then find the solution by tuning $r_a$ for the demanding asymptotic behavior, and the perturbed size $r'_H>r_H$ is found from the numerical fit.

\begin{figure}[!h]
  \centering%
{ \includegraphics[width=7.45cm]{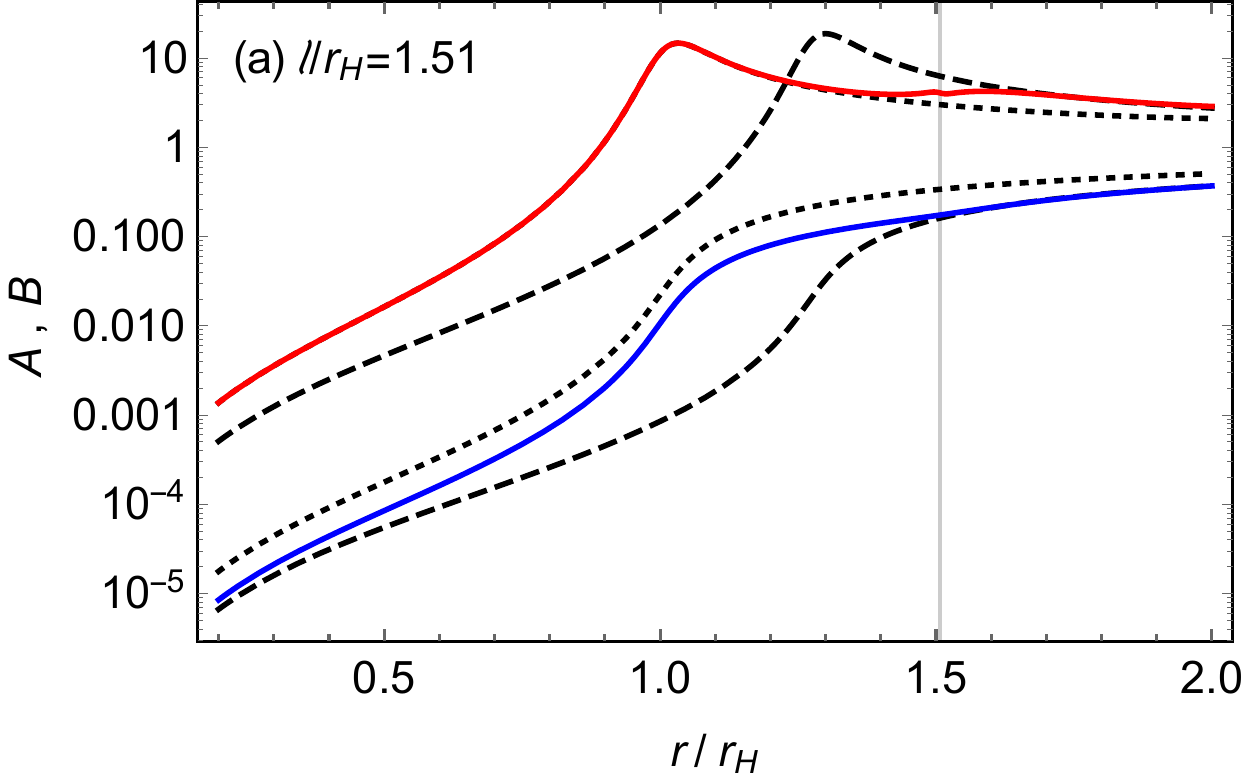}}\quad
{ \includegraphics[width=7.45cm]{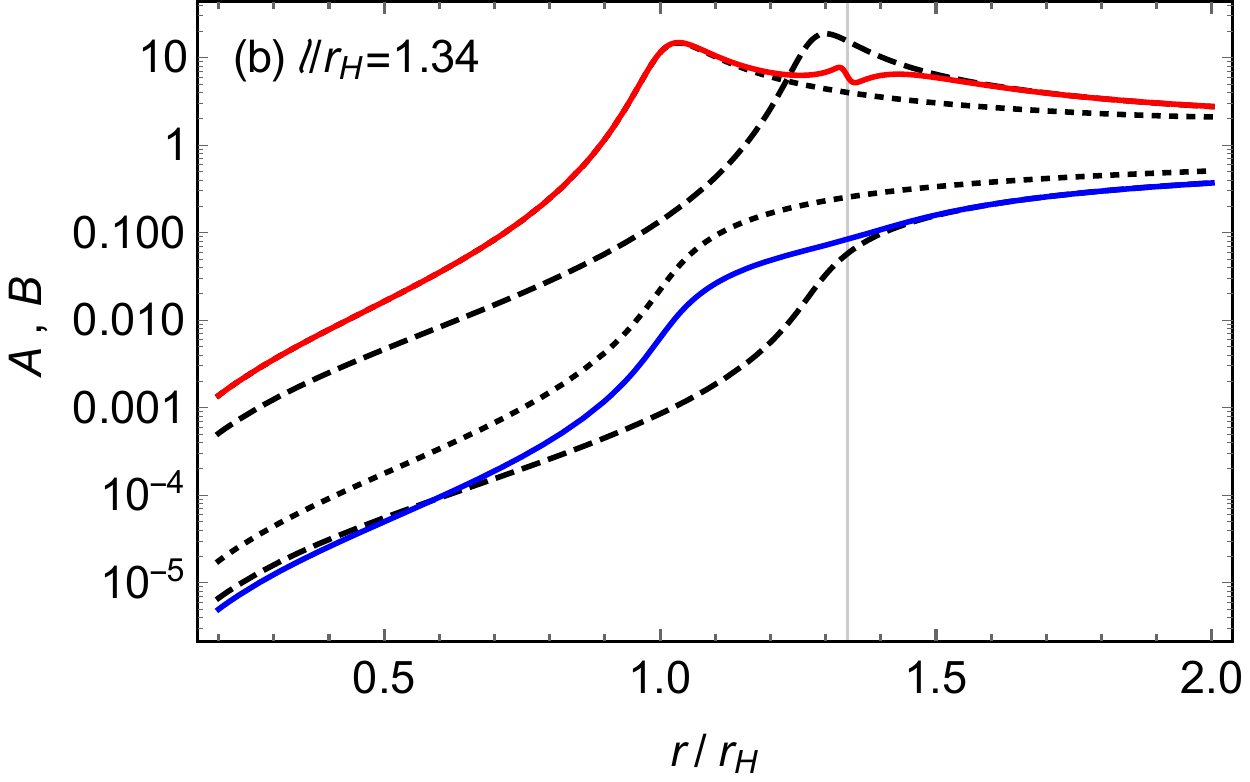}}\\
{ \includegraphics[width=7.45cm]{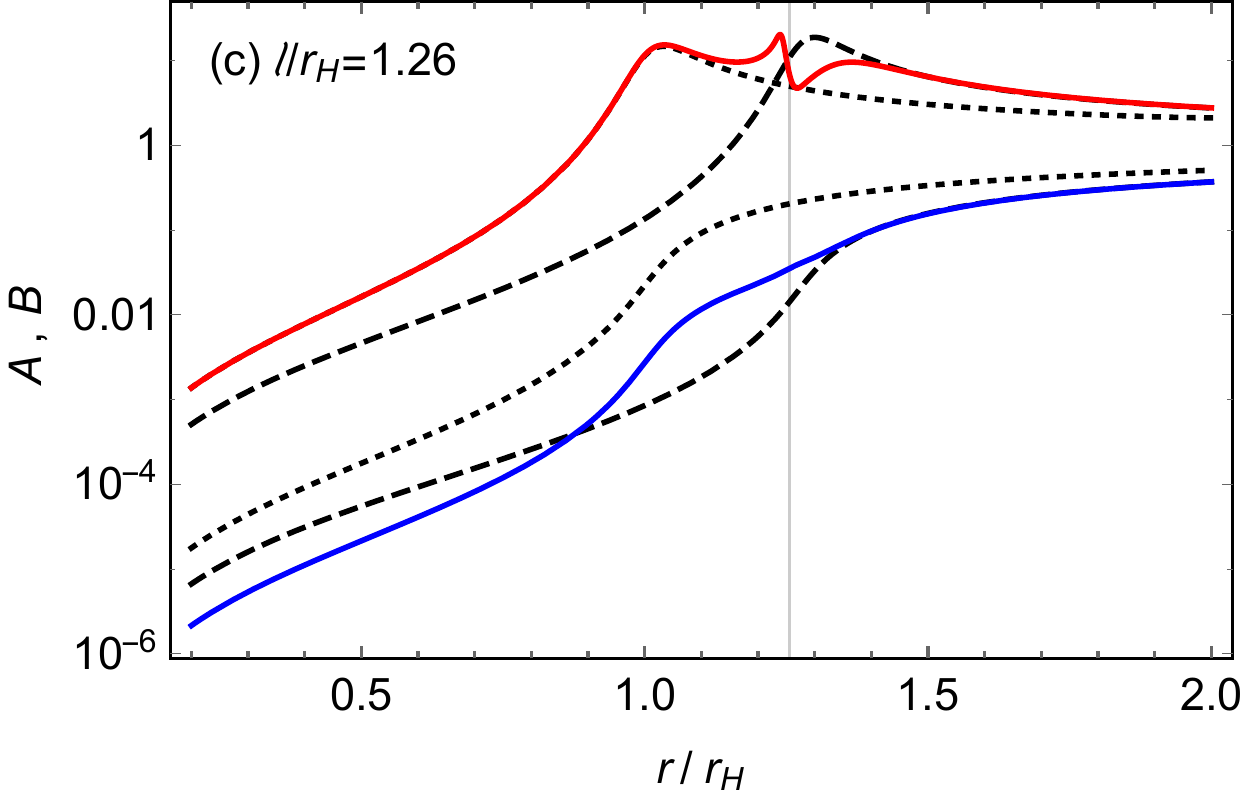}}\quad
{ \includegraphics[width=7.45cm]{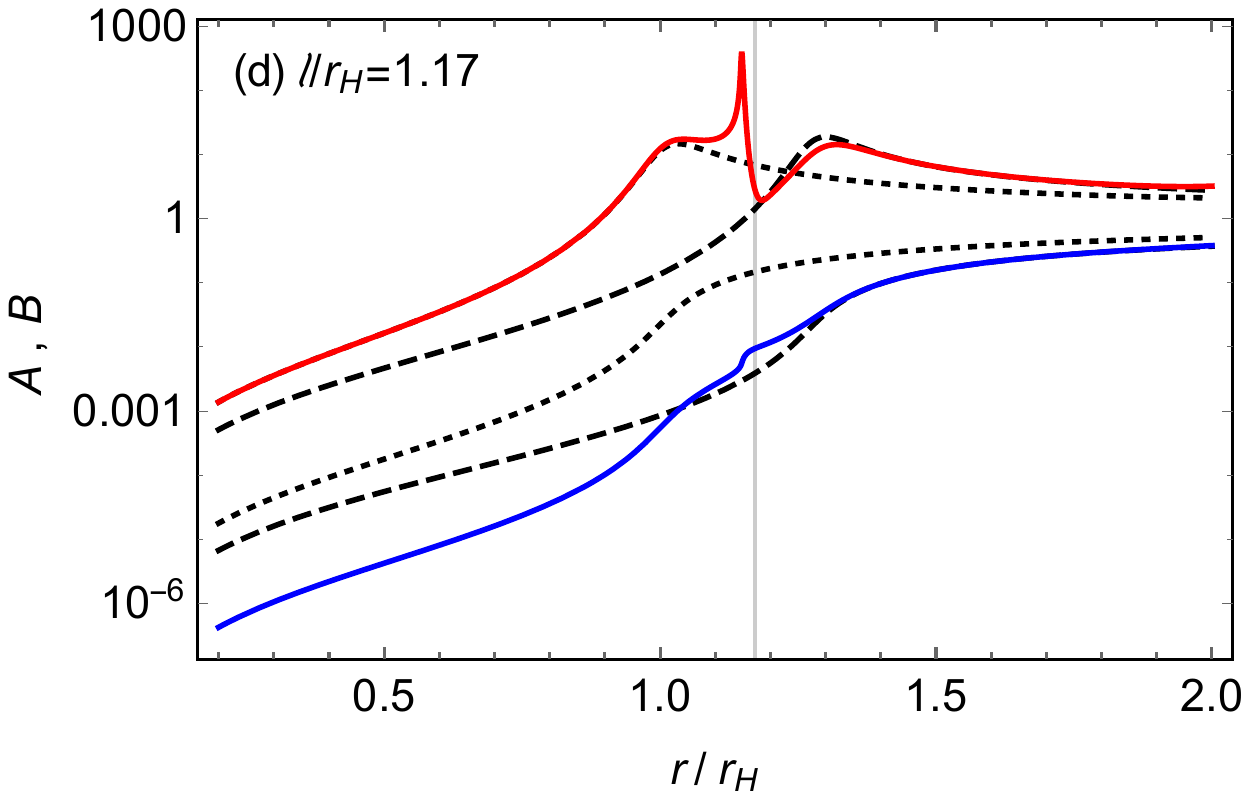}}\\
{ \includegraphics[width=7.45cm]{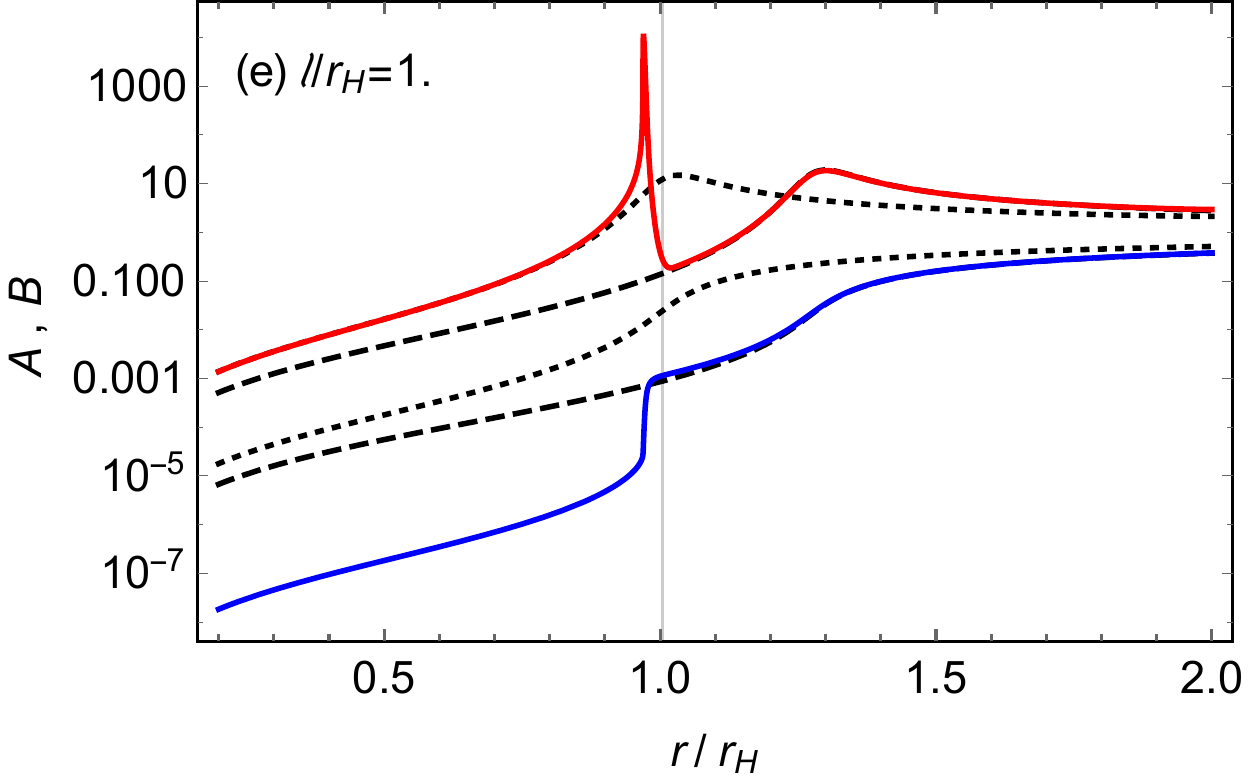}}\quad
{ \includegraphics[width=7.45cm]{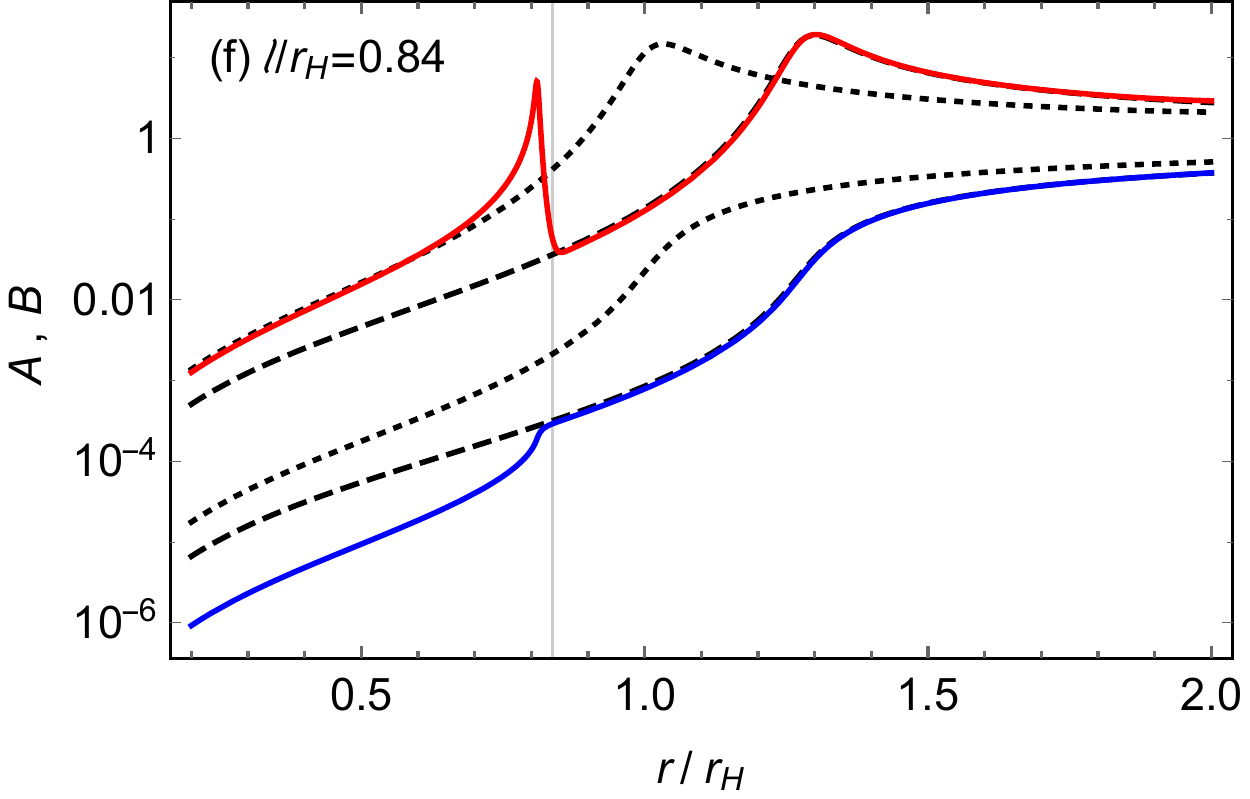}}
\caption{\label{fig:pert22h} 
$A(r)$ (red) and $B(r)$ (blue) for the 2-2-hole perturbed by a matter shell at different $\ell/r_H$ (vertical grey line), with $r'_H/r_H=1.27$ and $r_H/\lambda_2=6$. The black dot and dash lines denote the unperturbed 2-2-holes with the would-be horizon $r_H$ and $r'_H$ respectively (top line for $A(r)$, bottom line for $B(r)$).}
\end{figure}

For a given pair $(r_H, r'_H)$, there is then a unique static perturbed 2-2-hole at each $\ell$. We use this to model a quasi-static process of a matter shell slowly falling into a large 2-2-hole. Fig.~\ref{fig:pert22h} shows $A(r), B(r)$ of perturbed 2-2-holes for decreasing $\ell/r_H$, in comparison to the original and new unperturbed 2-2-holes with the would-be horizon $r_H$ and $r'_H$ respectively. For illustration we choose a large shell perturbation with the fraction $(r'_H-r_H)/r_H$ a few ten percent. For realistic cases, the fraction could be much smaller and the transition region is much narrower. When $\ell\gtrsim r'_H$, we see relatively small impact from the matter shell. The transition region agrees with the original 2-2-hole around $r_H$, while at large radius it is well approximated by the Schwarzschild metric with $r'_H$. When $r_H\lesssim \ell \lesssim r'_H$,  the transition region is significantly deformed by the matter shell. We see deviation from the original 2-2-hole around $r_H$ and recovery of the new 2-2-hole around $r'_H$. When $\ell\lesssim r_H$, the solution becomes indistinguishable from the new 2-2-hole outside the shell, while it is more similar to the original 2-2-hole inside the shell. 
With the shell moving inwards, the scales related to the interior $r_a, r_b$ also change.

This shows explicitly the response of a 2-2-hole to a matter shell, in particular the change of the transition region. 
In some toy models of horizonless ultracompact objects, the transition region is simply assumed to be a matter surface close to the would-be horizon. Given the causality constraint on the growth rate of the matter surface, an upper limit on the compactness of horizonless object is derived in \cite{Carballo-Rubio:2018vin}. Basically, if matter accretion proceeds too fast, the surface expansion in the large redshift region cannot catch up with the growth rate of the would-be horizon, and the object will turn to a black hole. 
The perturbed 2-2-hole as shown in Fig.~\ref{fig:pert22h} provides a counter-example for such arguments. Instead of an expanding surface, the transition region of the 2-2-hole varies in a rather complicated way for a series of $\ell$. Since the thermal gas density drops dramatically outside $r_H$, the change is mainly for the background spacetime, which shall be free from the causality constraint. 
As is well known in cosmology, the expansion of universe could safely be superluminal.

The matter shell perturbation for fixed $(r_H, r'_H)$ has similar properties to the thin-shell model~\cite{Holdom:2016nek}. The equation of state $p_\ell/\rho_\ell$, as determined by the momentum conservation law $\ell B'(\ell)/4B(\ell)$, increases with respect to $\ell/r_H$ and violates the dominant energy condition $|p_\ell|\leq \rho_\ell$ at some intermediate radius. It reaches a maximum around $\ell\approx r_H$ and then declines as for the Schwarzschild metric. If the shell moves from one radius to another with time-dependent $\ell$, we have also checked that the stress tensor satisfies the energy conservation law $\nabla^\mu T_{\mu t}=0$ \footnote{The law is $d\sigma_\ell/d\ell+2\sigma_\ell/\ell\left(1+\ell B'(\ell)/4B(\ell)\right)=0$ for the properly normalized density $\sigma_\ell\equiv \rho_\ell(\ell)\sqrt{A(\ell)}$.} within numerical errors. 

Given the special equation of state for the matter shell, we cannot directly calculate its entropy. However it shall be smaller than entropy of the relativistic thermal gas with the same energy density, i.e. $s(r)\leq 4/3(3N/\pi^2)^{1/4}\rho(r)^{3/4}$. This can provide a loose upper bound on the total entropy for the perturbed 2-2-hole. With decreasing $\ell/r_H$, the thermal gas contribution declines from the original 2-2-hole value due to the shell's backreaction on the spacetime, while the maximal contribution from the shell grows. The upper bound for the total entropy $S_{\textrm{max}}$ turns out to increase, with $S_{\textrm{max}}<S'$ at $\ell/r_H\gtrsim 1$ and $S_{\textrm{max}}>S'$ at some small $\ell$, where $S'$ denotes entropy of the new unperturbed 2-2-hole. 
Notice that we ignore interaction between the matter shell and the relativistic thermal gas in above discussion. In a more physical scenario, the interaction may become significant when the shell falls into the 2-2-hole interior filled with the high temperature gas. The shell then gets burned up and reaches thermal equilibrium with the gas after some time. The configuration ends up to approach the new 2-2-hole. This process shall respect the generalized second law of thermodynamics, and $S_{\textrm{max}}<S'$ for $\ell/r_H\gtrsim 1$ is expected.\footnote{In the limit of negligible interaction, $S_{\textrm{max}}>S'$ for the shell deep inside implies an upper bound for its entropy.}

\section{Conclusion}
\label{sec:conc}

The new era of observational astronomy provides a great opportunity to test horizonless ultracompact objects as black hole mimickers. Among all, the 2-2-hole in quadratic gravity is an interesting candidate. 
In this paper, we drew an overall picture for 2-2-holes as sourced by the thermal gas, which might more appropriately describe the final form of infalling matter during gravitational collapse. The metric and matter properties in the relativistic thermal gas model are illustrated in Fig.~\ref{fig:m2rH1}. The essential features are captured by the large mass and small mass limits, with different scaling behaviors summarized in Tab.~\ref{tab:scaling}. 

As departures from black holes are restricted to be small in the exterior, 
astrophysical 2-2-holes are probably in the large mass limit $r_H/\lambda_2\gg1$. Their properties are determined by the macroscopic size $r_H$ as well as the microscopic scale $\lambda_2$. Black hole thermodynamics, such as the area law for entropy and the inverse mass dependence for temperature in (\ref{eq:TSLmass}), becomes an emergent phenomenon in this limit. The area law in particular results from the large hierarchy between the small radial proper length $\sim\lambda_2$ and the large angular one $\sim r_H$ for the interior. 
The numerical values of thermodynamic variables are nonetheless different. A 2-2-hole can easily be entropically preferred over a comparable black hole and may serve as the endpoint of gravitational collapse.
As a reference value, $r_H/\lp\sim 10^{40}$ for stellar mass objects. So the astrophysical 2-2-hole interior literally approaches a firewall of negligible width, which is filled with the extremely hot gas. The singularity is sitting almost right at the would-be horizon. 
When the size of 2-2-holes becomes comparable to $\lambda_2$ with $M\sim \Mp^2/m_2$, the behavior is governed by the other length scale $r_a$ $(\ll \lambda_2)$, which describes its shrinking interior (and also the normalization scale $r_b$). As a result, the entropy scales more like what we may naively expect for the self-gravitating radiation inside a box, and the heat capacity turns positive as in (\ref{eq:TSSmass}). The departure from black hole thermodynamics suggests that a small 2-2-hole will become colder and radiate slower at the later stage of evaporation. Instead of an explosion, a burning 2-2-hole ends up with a small remnant with $r_H\sim \lambda_2$, which may naturally serve as dark matter.  
The first law of thermodynamics is realized differently in the two limits, with quite distinct energy budgets. 

Note that the main discussion of 2-2-hole thermodynamics here is in the context of classical physics. It has nothing to do with quantum effects or specialties of the horizon as for the black hole. 
The fact that black hole thermodynamics emerges from that of ordinary matter at certain limit 
may shed a fresh light on the relation between geometry and thermodynamics. For example, it would be interesting to study the relevance of the entropy bound and the holographic principle in the context of horizonless objects.  

As variations of the simplest model, we explored the impact of the gas particle mass on the solution. A large mass can significantly change the interior matter distribution as in Fig.~\ref{fig:mass}, but the explicit values of thermodynamic variables such as gas temperature, entropy and energy are found to be quite insensitive to such details of matter. 
So the formation of a highly curved but horizonless region scrambles initial information of infalling matter to some extent and makes it less accessible from the outside. Unlike the black hole, there is no information loss.  
We also studied a series of configurations with a matter shell perturbation at different radii, as snapshots of a 2-2-hole with a slow accretion of matter.
Fig.~\ref{fig:pert22h} provides an explicit picture for how the 2-2-hole spacetime (in particular the transition region) expands. In contrast to a hard surface as employed in some toy models, this is not subjected to the causality constraint. And there is no need to form a black hole.

We haven't checked the radial stability for the thermal gas model in this work. Given that the 2-2-hole mass increases monotonically with the gas central density (say the $1/r^4$ term coefficient), it is tempting to speculate that all 2-2-holes stay at the same stability branch. Nonetheless, the commonly used variation principle for stability might not apply here. The characteristic frequencies of oscillation need to be checked explicitly. The instability as associated with the classical ghost mode on the other hand might be an artifact of the classical approximation, and shall be accounted for by quantum corrections in the full theory. 
The study of stability and dynamics is crucial for the question about the endpoint of a gravitational collapse in quadratic gravity.
There are interesting phenomenological implications to explore. For gravitational wave echoes, the thermal gas in the 2-2-hole interior can lead to the damping of gravitational wave. 
This may provide a benchmark for the echo study, and possibly a resolution to the ergoregion instability for rotating horizonless objects. 
For the dark matter physics, primordial 2-2-holes in the small mass limit have very different thermodynamic behaviors. If they are the dominant dark matter constituent, current observational constraints can be changed and a new window may open.

\begin{acknowledgments}
We would like to thank Bob Holdom for early collaboration and valuable discussions. 
This research is supported in part by the Institute of High Energy Physics under Contract No. Y8515560U1. 

\end{acknowledgments}

\appendix

\section{Field equations in the Einstein-Weyl theory}
\label{sec:appA}

The field equation in the Einstein-Weyl theory is~\cite{Lu:2015psa},
\begin{eqnarray}\label{eq:EoMCV}
H_{\mu\nu}&\equiv&
\Mp^2\left(R_{\mu\nu}-\frac{1}{2}g_{\mu\nu}R\right)
-4\alpha \, B_{\mu\nu}=8\pi T_{\mu\nu}\,.
\end{eqnarray}
$B_{\mu\nu}$ is the traceless and symmetric Bach tensor,
\begin{eqnarray}
B_{\mu\nu}&\equiv&\left(\nabla^\rho\nabla^\sigma+\frac{1}{2}R^{\rho\sigma}\right)C_{\mu\rho\nu\sigma}\nonumber\\
&=&R_{\mu \rho \nu \sigma } R^{\rho \sigma }-\frac{1}{6}\nabla _{\mu }\nabla _{\nu } \,R
+\frac{1}{2}\Box \, R_{\mu \nu }
-\frac{1}{12}g_{\mu \nu }\Box\, R
-\frac{1}{3}R\, R_{\mu \nu }
-\frac{1}{4}g_{\mu \nu } \left[R^{\rho \sigma } R_{\rho \sigma }-\frac{R^2}{3}\right]
\end{eqnarray}

Since $H_{\mu\nu}$ satisfies the Bianchi identity, there are only two independent equations from (\ref{eq:EoMCV}) for a static and spherically symmetric spacetime (\ref{eq:ds2}). And we choose the following two combinations,
\begin{eqnarray}\label{eq:EoM1}
H_1=8\pi T_\mu^\mu,\quad H_2=8\pi T_2\,.
\end{eqnarray}
The first equation is the trace of (\ref{eq:EoMCV}). It depends only on the Einstein term and trace of the stress tensor,
\begin{eqnarray}
H_1&=&-\Mp^2 R\nonumber\\
&=&\frac{-\Mp^2}{r^2 A^2 \left(r B'-2 B\right)}\Big[r B A' \left(r B'+4 B\right)+A \left(r^2 B'^2-2 B \left(r^2 B''+2 r B'\right)-4 B^2\right)+4 A^2 B^2\Big].\nonumber\\
\end{eqnarray}
The derivatives are all with respect to $r$. $H_2$ includes the essential contribution from the Weyl tensor term,
\begin{eqnarray}
H_2&=&\frac{\Mp^2}{r^2B}(B+r B'-A B)
+\frac{\Mp^2 \lambda_2^2}{4 r^4 A^3 B^3}\Big[r^2 B^2 A'^2 \left(5 B-4 r B'\right)+A^2 \left(r^3 B'^3-3 r^2 B B'^2-4 B^3 \left(r A'+2\right)\right)\nonumber\\
&&+A B \left(r^3 A' B'^2+2 r B B' \left(r^2 A''+r A'\right)+4 B^2 \left(r A'-r^2 A''\right)\right)+8  A^3 B^3\Big],
\end{eqnarray} 
where $m^2_2=1/\lambda_2^2=\Mp^2/2\alpha$. The stress tensor is also more complicated
\begin{eqnarray}
T_2&=&T_{rr}
-X \frac{2B^2}{r B'-2B}T_\mu^\mu
-Y r \left(\frac{2B^2}{r B'-2B}T_\mu^\mu\right)',\nonumber\\
X&=&\frac{r B'-2 B}{48 A B^4}\frac{\lambda_2^2}{r^2} \left[r B A' \left(r B'-8 B\right)+A \left(4 B^2-7 r^2 B'^2+2 B \left(r^2 B''+8 r B'\right)\right)-4 A^2 B^2\right],\nonumber\\ 
Y&=&\frac{(r B'-2B)^2}{12 B^3}\frac{\lambda_2^2}{r^2}\, .
\end{eqnarray} 
$H_1$ is of second differential order in $B(r)$ and first order in $A(r)$, while $H_2$ is of second order in $A(r)$ and first order in $B(r)$. Such a symmetric pattern makes it easier to find numerical solutions~\cite{Lu:2015psa}.

With the spherical symmetry, a general stress tensor for matter is
\begin{eqnarray}\label{eq:stresstensorG}
T_{\mu\nu}=\textrm{diag}\,\left(B \rho, \,A\, p_r, \,r^2 p_\theta, \,r^2 s_\theta^2 p_\theta\right),
\end{eqnarray} 
where $\rho, p_r, p_\theta$ are proper energy density and pressure. The stress tensor satisfies the momentum conservation law $\nabla^\mu T_{\mu r}=0$,
\begin{eqnarray}\label{eq:Mcons}
p'_r+\left(\frac{2}{r}+\frac{B'}{2B}\right)p_r+\frac{B'}{2B}\rho-\frac{2}{r}p_\theta=0\,.
\end{eqnarray}
To solve the whole system, two more equations of state are needed to specify the relation of $\rho, p_r, p_\theta$. The conservation law is of first differential order in matter properties.

For a given matter source, we then combine two field equations (\ref{eq:EoM1}) and the conservation law (\ref{eq:Mcons}) to find $A(r), B(r), F(r)$, where $F(r)$ denotes one matter function after the other two eliminated by equations of states. 
Note that all equations are independent of a constant rescaling of $B(r)$ (or equivalently a rescaling of $t$), corresponding to $b_2$ in the series expansion. For asymptotically flat solutions, this rescaling is fixed by the normalization $B(\infty)=1$. So in general the system can be solved with four initial conditions $\{A', A, B'/B, F\}$ at some $r=r_0$.

\section{Series expansion for the thermal gas model}
\label{sec:appB}

The solutions in classical quadratic gravity can be classified by the series expansion around the origin $r=0$,
\begin{eqnarray}
A(r)&=&a_s r^s+a_{s+1}r^{s+1}+a_{s+2}r^{s+2}+...\,,\nonumber\\
B(r)&=&b_t (r^t+b_{t+1}r^{t+1}+b_{t+2}r^{t+2}+...)\,.
\end{eqnarray}
There are three families of solutions as characterized by the powers of the first nonvanishing terms $(s,t)$ \cite{Stelle:1977ry}. The $(0,0)$ family is nonsingular at the origin. The $(1,-1)$ family includes the Schwarzschild solution as a special case. The $(2,2)$ family with vanishing metric at the origin is a new type of solution and also the most generic family in quadratic gravity~\cite{Holdom:2002xy}.

Here we focus on a subclass of 2-2-holes that has only even power terms in the series expansion. 
For the thermal gas model, the series expansion is governed by four quantities: the Compton wavelength for the new spin-2 mode $\lambda_2$, the gas particle mass $m$, the new scale $r_a=1/\sqrt{a_2}$ and the coefficient for the gas temperature $c_{-1}$. We list the first few terms in the following
\begin{eqnarray}\label{eq:22SE}
A(r)&=& \frac{r^2}{r_a^2}\left[1+\frac{r^2}{r_a^2}\left(\frac{4}{3}\sqrt{1+9\frac{r_a^2}{\lambda_2^2}-\frac{3}{2\pi^2}\frac{\lp^2}{{\lambda_2^2}}c_{-1}^4}-\frac{1}{3}+\frac{\lp^2 m^2}{24\pi^2}c_{-1}^2\right)+\mathcal{O}(r^4)\right], \nonumber\\
B(r)&=&\frac{r^2}{r_b^2}\left[1+\frac{r^2}{r_a^2}\sqrt{1+9\frac{r_a^2}{\lambda_2^2}-\frac{3}{2\pi^2}\frac{\lp^2}{{\lambda_2^2}}c_{-1}^4}+\mathcal{O}(r^4)\right], \nonumber\\
T(r)&=&\frac{c_{-1}}{r}\left[1-\frac{r^2}{2r_a^2}\sqrt{1+9\frac{r_a^2}{\lambda_2^2}-\frac{3}{2\pi^2}\frac{\lp^2}{{\lambda_2^2}}c_{-1}^4}+\mathcal{O}(r^4)\right].
\end{eqnarray} 
The parameter $r_b=1/\sqrt{b_2}$, as related to the normalization of $B(r)$, is determined by matching with the asymptotically flat solution at large radius with $B(\infty)=1$. 
We can see that the particle mass only enters at the subleading order. In the massless limit, we have
\begin{eqnarray}\label{eq:22SEM0}
A(r)&=& \frac{r^2}{r_a^2}\left[1
+\frac{r^2}{r_a^2}\left(\frac{4}{3}\mathcal{D}-\frac{1}{3}\right)
+\frac{r^4}{r_a^4}\left(\frac{3}{2}+\frac{27r_a^2}{2\lambda_2^2}-\frac{5}{6}\mathcal{D}-\frac{31}{12\pi^2}\frac{\lp^2}{{\lambda_2^2}}c_{-1}^4\right)
+\mathcal{O}(r^6)\right], \nonumber\\
B(r)&=&\frac{r^2}{r_b^2}\left[1
+\frac{r^2}{r_a^2}\mathcal{D}
+\frac{r^4}{r_a^4}\left(\frac{7}{9}+\frac{15r_a^2}{2\lambda_2^2}-\frac{1}{9}\mathcal{D}-\frac{17}{12\pi^2}\frac{\lp^2}{{\lambda_2^2}}c_{-1}^4\right)
+\mathcal{O}(r^6)\right], \nonumber\\
T(r)&=&\frac{c_{-1}}{r}\left[1
-\frac{r^2}{2r_a^2}\mathcal{D}
-\frac{r^4}{r_a^4}\left(\frac{1}{72}+\frac{3r_a^2}{8\lambda_2^2}-\frac{1}{18}\mathcal{D}-\frac{7}{48\pi^2}\frac{\lp^2}{{\lambda_2^2}}c_{-1}^4\right)
+\mathcal{O}(r^6)\right].
\end{eqnarray} 
where $\mathcal{D}=(1+9r_a^2/\lambda_2^2-3\lp^2c_{-1}^4/(2\pi^2\lambda_2^2))^{1/2}$. The last two expressions satisfy the conservation law: $T(r)B(r)^{1/2}=c_{-1}/r_b+\mathcal{O}(r^6)$.

In the large mass limit, (\ref{eq:22SE}) can be simplified by only keeping the leading order terms for $r_a/\lambda_2\gg1$
\begin{eqnarray}\label{eq:22SELM}
\frac{r_H^2}{\lambda^2_2}A(r)&=& \frac{1}{\tilde{r}_a}\Bigg[\tilde{r}^2
+4\tilde{r}^4\left(\sqrt{1-\frac{\tilde{c}_{-1}^4}{6\pi^2}}+\frac{\tilde{c}_{-1}^2\tilde{m}^2}{96\pi^2}\right)
+\tilde{r}^6\Bigg(\frac{27}{2}-\frac{31\tilde{c}_{-1}^4}{12\pi^2}+\frac{\tilde{c}_{-1}^2\tilde{m}^2}{4\pi^2}\sqrt{1-\frac{\tilde{c}_{-1}^4}{6\pi^2}}\nonumber\\
&&+\frac{\tilde{c}_{-1}^4\tilde{m}^4}{576\pi^4}+\frac{\tilde{m}^4}{96\pi^2}\bigg(\frac{3}{2}+\gamma+\ln\frac{\tilde{m}\tilde{r}}{2\tilde{c}_{-1}}\bigg)\Bigg)
+\mathcal{O}(\tilde{r}^8)\Bigg], \nonumber\\
\frac{r_b^2}{r_a^2}\frac{r_H^2}{\lambda^2_2}B(r)&=&\frac{1}{\tilde{r}_a}\left[\tilde{r}^2
+3\tilde{r}^4\sqrt{1-\frac{\tilde{c}_{-1}^4}{6\pi^2}}
+\tilde{r}^6\left(\frac{15}{2}-\frac{17\tilde{c}_{-1}^4}{12\pi^2}+\frac{\tilde{c}_{-1}^2\tilde{m}^2}{24\pi^2}\sqrt{1-\frac{\tilde{c}_{-1}^4}{6\pi^2}}+\frac{\tilde{m}^4}{96\pi^2}\right)+\mathcal{O}(\tilde{r}^8)\right], \nonumber\\
(\lambda_2\lp)^{\frac{1}{2}}T(r)&=&\tilde{c}_{-1}\Bigg[\frac{1}{\tilde{r}}
-\frac{3}{2}\tilde{r}\sqrt{1-\frac{\tilde{c}_{-1}^4}{6\pi^2}}
-\tilde{r}^3\left(\frac{8}{3}-\frac{7\tilde{c}_{-1}^4}{48\pi^2}+\frac{\tilde{c}_{-1}^2\tilde{m}^2}{48\pi^2}\sqrt{1-\frac{\tilde{c}_{-1}^4}{6\pi^2}}+\frac{\tilde{m}^4}{192\pi^2}\right)\nonumber\\
&&+\mathcal{O}(\tilde{r}^5)\Bigg],
\end{eqnarray} 
where $\gamma$ is the Euler constant. According the scaling behavior in the large mass limit in Tab.~\ref{tab:scaling}, we rearrange various scales into dimensionless quantities as
$\tilde{r}_{a}\equiv r_{a} \lambda_2/r_H^{2}$, $\tilde{r}\equiv r/\sqrt{r_a\lambda_2}$, $\tilde{c}_{-1}\equiv c_{-1}\sqrt{\lp/r_a}$, $\tilde{m}\equiv m\,(\lambda_2\lp)^{1/2}$. The metric and temperature are then expansions of the small parameter $\tilde{r}$, and agree at the leading order at different $r_H/\lambda_2\gg1$ for a given set of $(\tilde{r}_{a}, \tilde{c}_{-1}, \tilde{m})$. 

In the small mass limit, we instead keep the leading order terms for $r_a/\lambda_2\ll1$ in (\ref{eq:22SE}),
\begin{eqnarray}\label{eq:22SESM}
A(r)&=& \tilde{r}^2
+\tilde{r}^4\left(1+\frac{\tilde{c}_{-1}^2\tilde{m}^2}{24\pi^2}\right)
+\tilde{r}^6\bigg[\frac{2}{3}+\frac{\tilde{c}_{-1}^2\tilde{m}^2}{18\pi^2}+\frac{\tilde{c}_{-1}^4\tilde{m}^4}{576\pi^4}\nonumber\\
&&+\frac{\tilde{m}^4}{96\pi^2}\bigg(\frac{3}{2}+\gamma+\ln\frac{\tilde{m}\tilde{r}}{2\tilde{c}_{-1}}\bigg)\bigg]
+\mathcal{O}(\tilde{r}^8), \nonumber\\
\frac{r_b^2}{r_a^2}B(r)&=&\tilde{r}^2
+\tilde{r}^4
+\tilde{r}^6\left(\frac{2}{3}+\frac{\tilde{c}_{-1}^2\tilde{m}^2}{72\pi^2}+\frac{\tilde{m}^4}{96\pi^2}\right)
+\mathcal{O}(\tilde{r}^8), \nonumber\\
(r_a\lp)^{\frac{1}{2}}T(r)&=&\tilde{c}_{-1}\left[\frac{1}{\tilde{r}}
-\frac{1}{2}\tilde{r}
+\tilde{r}^3\left(\frac{1}{24}-\frac{\tilde{c}_{-1}^2\tilde{m}^2}{144\pi^2}-\frac{\tilde{m}^4}{192\pi^2}\right)
+\mathcal{O}(\tilde{r}^5)\right].
\end{eqnarray} 
Similarly we rearrange various scales into dimensionless quantities as $\tilde{r}\equiv r/r_a$, $\tilde{c}_{-1}\equiv c_{-1}\sqrt{\lp/r_a}$, $\tilde{m}\equiv m\,(r_a\lp)^{1/2}$, according the scaling behavior in the small mass limit in Tab.~\ref{tab:scaling}. Note the different definitions of $\tilde{r}, \tilde{m}$ in this case. It is simpler than (\ref{eq:22SELM}) since the $r_H, \lambda_2$ dependence is now reduced to $r_a$ dependence.

\section{Anisotropic fluid model}
\label{sec:appC}

Under the spherical symmetry, a general stress tensor might be anisotropic as in (\ref{eq:stresstensorG}).
The momentum conservation law is then,
\begin{eqnarray}
p'_r+\left(\frac{2}{r}+\frac{B'}{2B}\right)p_r+\frac{B'}{2B}\rho-\frac{2}{r}p_\theta=0
\end{eqnarray}
Around the origin, with $B'/2B\propto 1/r$, the conservation law at the leading order is: $p'_r+4p_r/r+(\rho-p_r-2p_\theta)/r=0$. So the traceless condition $\rho-p_r-2p_\theta=0$ directly gives rise to $p_r\propto 1/r^4$, the same leading order behavior as the thermal gas model. Implementing the traceless condition, the conservation law becomes,
\begin{eqnarray}
p'_r+\left[\frac{2}{r}(1-a)+\frac{B'}{B}(1+a)\right]p_r=0
\end{eqnarray}
where  $p_\theta/p_r=a$ is some function of $r$. For $a(r)$ smaller (larger) than 1, the matter distribution is less (more) compact than the relativistic thermal gas. 

The complex scalar theory with $\mathcal{L}_\phi=\frac{1}{2}\partial^\mu\phi^\dag\partial_\mu\phi+V(\phi)$~\cite{revBS2} provides a concrete realization of the anisotropic stress tensor.  Assuming $\phi=e^{-i\omega t}\Phi(r)$, the stress tensor has  
\begin{eqnarray}
\rho&=&\frac{1}{2}\left(B^{-1}\omega^2\Phi^2+A^{-1}\Phi'^2+V\right), \nonumber\\
p_r&=&\frac{1}{2}\left(B^{-1}\omega^2\Phi^2+A^{-1}\Phi'^2-V\right)=\rho-V, \nonumber\\
p_\theta&=&\frac{1}{2}\left(B^{-1}\omega^2\Phi^2-A^{-1}\Phi'^2-V\right)=p_r-A^{-1}\Phi'^2,
\end{eqnarray} 
where anisotropic pressure is related to the spatial variation of the scalar field $\Phi'$. The momentum conservation law of stress tensor is equivalent to the scalar field equation of motion,
\begin{eqnarray}
\Phi''+\Phi'\left[\frac{1}{2}\left(\frac{B'}{B}-\frac{A'}{A}\right)+\frac{2}{r}\right]+\Phi\left(\frac{A}{B}\omega^2-A\frac{\partial V}{\partial \Phi^2}\right)=0
\end{eqnarray} 
Around the origin, it becomes $\Phi''+2\Phi'/r=0$ at the leading order,
and so $\Phi=c_1+c_2/r$. The solution $\Phi\propto 1/r$ implies $\rho\approx p_r\approx -p_\theta\approx \frac{1}{2}A^{-1}\Phi'^2\propto 1/r^6$. It turns out to be too singular and $A(r), B(r)\sim r^2$ is no longer a solution. So only $\Phi\propto r^0$ is allowed, with $\rho\approx p_r\approx p_\theta\approx \frac{1}{2}B^{-1}\omega^2\Phi^2\propto 1/r^2$. For this more regular case, we haven't found 2-2-hole solutions with the right transition region and asymptotic behavior by the shooting method. The lack of solution for the complex scalar field might be related to the fact that both field equations receive a dominant contribution from the nonzero trace. 


\linespread{1}

\end{document}